\documentclass[3p,number,sort&compress]{elsarticle}
\usepackage{titlesec}
\usepackage[colorlinks=true]{hyperref}
\usepackage{parskip}
\usepackage[utf8]{inputenc}
\usepackage{verbatim}
\usepackage[skip=5pt]{caption}
\usepackage{tikz}
\usepackage{siunitx}

\usepackage{lineno}

\usepackage{algorithm}
\usepackage{algpseudocode}

\usepackage{adjustbox}
\usepackage{array}

\newcommand{\LineIf}[2]{\State\algorithmicif\ {#1}\ \algorithmicthen\ {#2}}

\newcommand{\Continue}{\textbf{repeat loop}}

\usepackage{fancyhdr}
\fancypagestyle{plain}{%
  \fancyhf{}
  \fancyhead[L]{\sc Barnes}
  \fancyhead[R]{\sc Parallel Priority-Flood}
  \cfoot{\thepage}
  
  }
\biboptions{square}

\newcommand{\todo}[1]{}

\hypersetup{
  pdfauthor   = {Richard Barnes (ORCID: 0000-0002-0204-6040)},
  pdftitle    = {Parallel Priority-Flood Depression Filling For Trillion Cell Digital Elevation Models On Desktops Or Clusters},
  pdfkeywords = {parallel computing, hydrology, geographic information system (GIS), pit filling, sink removal},
  pdfproducer = {LaTeX},
  pdfcreator  = {pdfLaTeX}
}

\makeatletter
\newenvironment{breakablealgorithm}
  {% \begin{breakablealgorithm}
   \vspace{0.5em}
   \begin{center}
     \refstepcounter{algorithm}% New algorithm
     \hrule height.8pt depth0pt \kern2pt% \@fs@pre for \@fs@ruled
     \renewcommand{\caption}[2][\relax]{% Make a new \caption
       {\raggedright\vspace{-0.6em}\textbf{\ALG@name~\thealgorithm} ##2\par}%
       \ifx\relax##1\relax % #1 is \relax
         \addcontentsline{loa}{algorithm}{\protect\numberline{\thealgorithm}##2}%
       \else % #1 is not \relax
         \addcontentsline{loa}{algorithm}{\protect\numberline{\thealgorithm}##1}%
       \fi
       \kern2pt\hrule\kern2pt
     }
  }{% \end{breakablealgorithm}
     \kern2pt\hrule\relax% \@fs@post for \@fs@ruled
   \end{center}
  }
\makeatother

\begin{document}
\thispagestyle{empty}

Cite as Barnes. ``Parallel Priority-Flood Depression Filling For Trillion Cell Digital Elevation Models On Desktops Or Clusters". Computers \& Geosciences. Vol 96, Nov 2016, pp 56--68. doi: ``10.1016/j.cageo.2016.07.001".

\begin{frontmatter}
  \title{\sc Parallel Priority-Flood Depression Filling \\ For Trillion Cell Digital Elevation Models \\ On Desktops Or Clusters}

  \author[rb]{Richard Barnes\corref{cor_rb}}
  \ead{richard.barnes@berkeley.edu}
  \address[rb]{Energy \& Resources Group, Berkeley, USA}
  \cortext[cor_rb]{Corresponding author. ORCID: 0000-0002-0204-6040}

  \begin{abstract} %~??? words
  \noindent Algorithms for extracting hydrologic features and properties from
  digital elevation models (DEMs) are challenged by large datasets, which often
  cannot fit within a computer's RAM. Depression filling is an important preconditioning step to many of these algorithms. Here, I present a new, linearly-scaling
  algorithm which parallelizes the Priority-Flood depression-filling algorithm by subdividing a DEM into tiles.
  Using a single-producer, multi-consumer design, the new algorithm works
  equally well on one core, multiple cores, or multiple machines and can take
  advantage of large memories or cope with small ones. Unlike previous
  algorithms, the new algorithm guarantees a fixed number of memory access and
  communication events per subdivision of the DEM. In comparison testing, this results in the new algorithm running generally faster while using fewer resources than previous algorithms. For moderately sized tiles,
  the algorithm exhibits $\sim60\%$ strong and weak scaling efficiencies up to 48 cores, and linear time scaling across datasets ranging over three orders of magnitude. The largest dataset on which I run the algorithm has 2~trillion
  ($2\cdot10^{12}$) cells. With 48 cores, processing required 4.8 hours
  wall-time (9.3 compute-days). This test is three orders of magnitude larger
  than any previously performed in the literature. Complete, well-commented
  source code and correctness tests are available for download from a
  repository.
  \end{abstract}

%Immediately after the abstract, provide a maximum of 5 keywords, avoiding general and plural terms and multiple concepts (avoid, for example, 'and', 'of'). Be sparing with abbreviations: only abbreviations firmly established in the field may be eligible. Please note that Keywords should NOT include words that already appear in the title of the manuscript. These keywords will be used for indexing purposes.
\end{frontmatter}

%The algorithm presented here is superior to a virtual tile approach since it can
%guarantee that each subdivision will be read and written to memory a fixed
%number of times, independent of the overall size of the dataset considered.

%Even if success is possible, such algorithms often run slowly. This has led to
%research on parallel algorithms and algorithms which explicitly manage memory.
%I argue here that current approaches do not scale well. Parallel approaches
%require many nodes and frequent communication between these nodes.
%Memory-managing algorithms suffer from poor access locality, forcing them to
%read and write subdivisions of the data numerous times.

\section{Introduction}
Digital elevation models (DEMs) are representations of terrain elevations above
or below a chosen zero elevation. Raster DEMs, in which the data are stored as
a rectangular array of floating-point or integer values, are widely used in
geospatial analysis for estimating a region's hydrologic and geomorphic
properties, including soil moisture, terrain stability, erosive potential,
rainfall retention, and stream power. Many algorithms for extracting these
properties require that, by following flow directions downhill from one cell to
another, it is always possible to reach the edge of the DEM.

Depressions (see \citet{Lindsay2015} for a typology) are inwardly-draining
regions of a DEM which have no outlet and, therefore, confound such algorithms.
Although depressions may be representative of natural terrain, such as in the
Prairie Pothole Region of the United States, they may also result from technical
issues in the DEM's collection and processing, such as from biased terrain
reflectance or conversions from floating-point to integer
precision.~\citep{Nardi2008} Note that depressions are distinct from pits, which
are single DEM cells whose neighbors all have a higher elevation.

Depressions may be dealt with by filling them in to the level of their lowest
outlet, as will be done here. Several authors have argued that this
approach produces inferior results compared to approaches which either solely
breach depression walls or combine breaching and filling.~\citep{Lindsay2015,
Garbrecht1998, Grimaldi2007, Lindsay2005, Danner2007} As a particularly
egregious example of a situation in which breaching would be better,
\citet{Metz2010} shows one river along which 92\% of cells were adjusted by
depression-filling. However, a DEM may be modified extensively without
compromising results, depending on the nature of the analysis being done.
Additionally, breaching and hybrid approaches continue to lag behind recent
developments in depression-filling, including the one described here, both in
terms of execution times and the size of the DEM it is possible to process.

For a given DEM $Z$, depression-filling, such as described by this paper,
produces a new DEM $W$ defined by the following criteria~\citep{Planchon2002}:
\begin{enumerate}
  \item The elevation of each cell of $W$ is greater than or equal to its
  corresponding cell in $Z$.
  \item For each cell $c$ of $W$, there is a path that leads from $c$ to the
  boundary by moving downwards by an amount of at least $\epsilon$ between any
  two cells on the path, where $\epsilon$ may be zero. Such a path is referred
  to as an $\epsilon$-descending path.
  \item $W$ is the lowest surface allowed by properties (1) and (2).
\end{enumerate}
This paper considers only the most common case wherein $\epsilon=0$. Setting
$\epsilon>0$ requires more complex methods than those described here.

DEMs have increased in resolution from 30--90\,m in the recent past to the
sub-meter resolutions becoming available today. Increasing resolution has led
to increased data sizes: current DEMs are on the order of gigabytes and
increasing, with billions of cells. Even in situations where only comparatively
low-resolution data is available, a DEM may cover large areas: 30\,m Shuttle
Radar Topography Mission (SRTM) elevation data has been released for 80\% of
Earth's landmass.~\citep{Farr2007} While computer processing and memory
performance have increased appreciably, development of algorithms suited to
efficiently manipulating large DEMs is on-going.

If a DEM can fit into the RAM of a single computer, several algorithms exist
which can efficiently perform depression-filling operations (see
\citet{Barnes2014pf} for a review and \citet{Zhou2015} for the latest work in
this area). If a DEM cannot fit into the RAM of a single computer, other
approaches are needed.

In this paper, I will argue that existing approaches are inefficient and do not
scale well. I will then present a new algorithm which overcomes the problems
identified. The new algorithm is able to efficiently fill depressions in DEMs
with more than a trillion cells and will work on both single-core machines and
supercomputers. The algorithm achieves this by subdividing not just the data,
but the problem itself: it is able to limit communication to a fixed number of
events per subdivision and I/O to a fixed number of events per DEM cell. The
algorithm may also offer efficiency advantages even if a DEM can fit entirely
into RAM.

\section{Background}
\label{sec:alt_algs}

\begin{table*}
\footnotesize
\centering
\begin{tabular}{llllllll}
Source                                 & Year & Cells           & Resolution & Dimensions          & Adjective             & Time (min) & Min/Cell                         \\ \hline % & DEM                           \\ 
This paper (RichDEM)                   & 2016 & $2\cdot10^{12}$ & 10\,m      & $\sim$1,291,715$^2$ & \textit{rather} large & 287        & $8\cdot10^{-9}$              \\ % & Amazon River Basin (ASTER)    \\ 
%\citet{Bai2015}               & 2015 & $7\cdot10^9$    & 30\,m      & ??                  & large                 & 816        & $1\cdot10^{-7}$              \\ % & Amazon River Basin (ASTER)    \\
\citet{Gomes2012}                      & 2012 & $3\cdot10^9$    & 30\,m      & 50,000\,x\,50,000   & huge                  & 58         & $1\cdot10^{-8}$                 \\ % & SRTM Region 3                 \\
\citet{Do2010}                         & 2010 & $2\cdot10^9$    & ??         & 36,002\,x\,54,002   & huge                  & 21         & $1\cdot10^{-8}$                 \\ % & Loire River, France           \\
\citet{Do2011}                         & 2011 & $2\cdot10^9$    & ??         & 36,002\,x\,54,002   & huge                  & ??         &                    \\ % & Loire Bretagne, France        \\
\citet{Yildirim2015} (TauDEM)          & 2015 & $2\cdot10^9$    & 10\,m      & 45,056\,x\,49,152   & large                 & ??         &                    \\ % & Chesapeake Bay (NED)          \\
\citet{Arge2003} (GRASS)               & 2003 & $1\cdot10^9$    & 10\,m      & 33,454\,x\,31,866   & massive               & 3720       & $3\cdot10^{-6}$                 \\ % & Washington State              \\
\citet{Lindsay2015} (Whitebox GAT)     & 2015 & $9\cdot10^8$    & 3\,arc-sec & 37,201\,x\,25,201   & massive               & 8.6        & $1\cdot10^{-8}$              \\ % & Nile Basin                       \\
\citet{Tesfa2011}                      & 2011 & $6\cdot10^8$    & ??         & 24,856\,x\,24,000   & large                 & 20         & $3\cdot10^{-8}$              \\ % & Boise River basin                          \\ %& 50\,s (C), 20 (T) \\
\citet{Wallis2009depressions} (TauDEM) & 2009 & $4\cdot10^8$    & ??         & 14,949\,x\,27,174   & large                 & 8          & $2\cdot10^{-8}$              \\ % & Great Salt Lake Region (NED)  \\
\citet{Danner2007}                     & 2007 & $3\cdot10^8$    & 3\,m       & ??                  & massive               & 445        & $1\cdot10^{-6}$              \\
\citet{Metz2010,Metz2011} (GRASS)      & 2010 & $2\cdot10^8$    & 30\,m      & ??                  & massive               & 32         & $6\cdot10^{-7}$              \\ \hline % & All of Panama                 \\
\end{tabular}
\caption{DEM sizes, dimensions, and processing times for authors working with
large DEMs. The table should be used only to develop a sense of the maximum
sizes and the range of times it can take to process large DEMs. Times between
algorithms should not be directly compared as different hardware has been used
in all cases and different operations have been performed in many cases. For
instance, \citet{Yildirim2015} performs depression-filling while
\citet{Lindsay2015} performs depression breaching. The authors' description of
the size of their data is also included; all authors used ``large". Some algorithms are part of larger terrain analysis suites, these are listed in parentheses.
\label{tbl:whatismassive}}
\end{table*}

Existing algorithms have taken one of two approaches to DEMs that cannot fit
entirely into RAM. They either (a)~keep only a subset of the DEM in RAM at any
time by using virtual tiles stored to a computer's hard disk or (b)~keep
the entire DEM in RAM by distributing it over multiple compute nodes which
communicate with each other. I argue here that existing algorithms pay high
costs in terms of disk access and/or communication which prevent them from
scaling well; the new algorithm pays much lower costs.

\autoref{tbl:whatismassive} lists several authors mentioned here who have
developed algorithms specifically for large DEMs. The sizes of the largest DEMs
they test are listed, along with their choice of adjective to describe this
size. Gigacell ($10^9$ cells) DEMs represent the upper limit of these tests.
Here, I will go further than ``massive" and bigger than ``huge" by testing a
trillion cell, or teracell ($10^{12}$ cells), DEM. After ruling out ``ginormous", I refer to this
new size class as being \textit{rather} large.

\subsection{Virtual Tiles}

The virtual tile approach subdivides a DEM into tiles, a limited number of
which can fit into RAM at a given time. When the RAM is full, tiles which are
not being used are written to the hard disk. Virtual tiles are advantageous
because they can be easily incorporated into any existing algorithm by
modifying the algorithm so that it accesses data through a tile manager. The
tile manager maps cells to tiles and, if the tile is not in memory, retrieves
it, possibly writing an old tile to disk first. Since hard disk access is slow,
existing algorithms reduce I/O by favoring access to nearby rather than distant
cells. This helps increase the locality of access, which is favourable for
caching. Unfortunately, virtual tile algorithms are unable to make strong
locality guarantees and therefore, are ultimately unable to limit how often a
particular tile will be loaded into memory.

\citet{Arge2003}, whose work is encapsulated in the
\textsc{TerraFlow}\footnote{\url{http://www.cs.duke.edu/geo*/terraflow/}}
package and included with GRASS~\citep{GRASS_GIS_software}, was one of the first
to examine I/O efficient algorithms for depression-filling (among other
operations). As discussed in their paper, since disk access is costly, blocks of
data are read from memory in an attempt to amortize this cost.
\citeauthor{Arge2003} describe a depression-filling algorithm which is bounded
by $O(N \log N)$ I/Os and $O(N\log N)$ operations (see their paper and
\citet{Aggarwal1988} for a more exact description of the access complexity).
Details of the algorithm's memory management are not described. They compared
the speed of their algorithm against ArcInfo 7.1.2 (an industry-standard for the
time) and achieved run-times twice as fast and completed larger problems.
\citet{Danner2007} describes an algorithm similar to \citet{Arge2003}, but
theirs performed a breaching operation on depressions.

\citet{Metz2011} present a Priority-Flood~\citep{Barnes2014pf} depression-breaching algorithm (now included with GRASS). The algorithm uses the GRASS segment library as a tile manager and, in comparison testing, achieves run-times almost twice as fast as \citet{Arge2003}, though the authors note that they expect that the algorithm by \citet{Arge2003} would be faster on larger datasets.

\citet{Gomes2012} present a virtual tile approach using an $O(N)$ integer
variant Priority-Flood in their EMFlow
package\footnote{\url{https://github.com/guipenaufv/EMFlow}}. The DEM is
subdivided into tiles accessed via a tile manager. Tiles are compressed before
being written to memory to limit the amount of memory to be written and, later,
read; this halves the execution time of the algorithm. Locality is achieved by
using a ``least recently used" (LRU) cache to evict the least-recently used tile
from memory. As the flood proceeds, it \textit{may} produce ``islands" of
unprocessed terrain. These islands, if present, are detected and processed one
at a time, which further increases locality. The algorithm out-performs that of
\citet{Arge2003} by $\sim$20\,x on the largest DEMs they consider. Their
implementation is limited to 2-byte integer data on square datasets.

\citet{Yildirim2015}
present\footnote{\url{https://bitbucket.org/ahmetartu/hydrovtmm}} a similar
algorithm with the addition of shared memory parallel processing. The input DEM
is divided into tiles and each tile is associated with its own thread. The
threads then perform some computations in parallel and regularly synchronize
their border information. Tiles are managed by a centralized thread which swaps
out the least recently used tile and tries to prefetch tiles it anticipates
will be needed. At its heart, the \citeauthor{Yildirim2015} algorithm relies on
the \citet{Planchon2002} algorithm, which repeatedly sweeps the entire DEM
until all depressions are filled.

Though some authors continue to base their work on the Planchon--Darboux
algorithm \citep{Yildirim2015,Yao2015} and many practitioners use it, there is
good evidence to suggest that it has been superseded by the Priority-Flood: for
their largest dataset, \citet{Wang2006} find that their variant of the
Priority-Flood algorithm runs 3\,x faster than \citeauthor{Planchon2002}. In
turn,
\citet{Barnes2014pf}\footnote{\url{https://github.com/r-barnes/Barnes2013
-Depressions}} achieve run-times 16\% faster than \citeauthor{Wang2006}. \citet{Zhou2015}\footnote{\url{https://github.com/zhouguiyun-uestc/FillDEM}}
achieve run-times 44.6\% faster than \citeauthor{Barnes2014pf}. These speed-ups
are due to continuous decreases in the time complexity of the algorithms
involved, from the $O(N^{1.2})$ complexity of the \citeauthor{Planchon2002}
algorithm to the $O(m \log m)$ with $m\ll N$ complexity of the
\citeauthor{Zhou2015} algorithm.

The algorithms above all use virtual tile methods to handle DEMs too
large to fit into RAM. Although a range of techniques are used to increase
access locality and speed including island detection, LRU-caches, and
prefetching, all of the underlying depression-filling algorithms work by
flooding terrain inwards from the perimeter of the DEM. Therefore, all of these
algorithms must at least load the entire perimeter before they can begin
flooding terrain. Walking the perimeter of a DEM is inherently non-local and,
for large DEMs with long perimeters, many tiles may be loaded and evicted from
the cache.

In general, there is no way to guarantee locality and disparate tiles may need
to be repeatedly loaded. Therefore, the number of memory accesses will tend to
increase with the size of the DEM; this decreases the scalability of the
algorithms, as will be demonstrated in \textsection\ref{sec:results}. The
algorithm presented here is superior to existing virtual tile approaches because
it can guarantee locality and ensure that each tile is accessed a fixed number
of times, regardless of the size of the DEM.

\subsection{Parallel, Multiple Nodes}

Distributing a DEM over several compute nodes may seem to be a solution to this
as the entire DEM can then be kept in RAM and, indeed, several authors have
pursued this path.

\citet{Wallis2009depressions} (part of
TauDEM\footnote{\url{https://github.com/dtarb/TauDEM}}) modify the aforementioned
\citet{Planchon2002} algorithm by dividing the DEM into a series of strips each
of which is managed by its own process. Each of the full DEM sweeps required by
\citeauthor{Planchon2002} is performed in parallel and all nodes communicate
with their neighbours after each sweep.

\citet{Do2010} and \citet{Do2011} calculate catchment basins and flow
accumulation, respectively, using a distributed minimum spanning tree algorithm.
Depressions are not explicitly treated. Their approach passes edge information
in addition to graph information between nodes, each of which holds a tile of
the larger DEM. They do not provide an analysis of their algorithm's
communication requirements nor source code. Using 8~processors their method is
approximately 10\,x faster than that of \citet{Arge2003}. Note that this implies
that their algorithm is out-performed by \citet{Gomes2012}.

\citet{Tesfa2011} assume a depression-filled DEM, divide the large DEM into
strips, and regularly synchronize information between the strips to calculate
hydrological proximity measures.

\citet{Yildirim2015}, as described above, uses parallel processing in shared
memory on a single machine to process a tiled DEM. This captures some of the
speed gains of a fully parallel approach while decreasing the number of
processors required by using a tile manager; however, their algorithm still
requires frequent interprocess communication.

%\citet{Bai2015} discuss a form of parallel processing wherein individual
%watersheds (or other hydrologic subdivisions) are apportioned to separate
%workers, claiming that it poses a complex problem which easily results in
%boundary errors and should, therefore, be avoided. They go on to develop a
%non-parallel approach based on the \citet{Wang2006} Priority-Flood modified to
%use a balanced binary search tree instead of a priority-queue. They claim the
%BST shows a 40\% improvement over the priority-queue, but it is unclear whether
%this is due to algorithmic efficiency or implementation differences. No source
%code is provided. Fortunately, the critique of \citet{Bai2015} is misplaced: it
%is possible to solve the depression filling problem without needing to
%explicitly identify hydrological subdivisions; as shown here, a tiled approach
%is both sufficient and efficient.

For large DEMs, strips such as those used by \citet{Wallis2009depressions} and
\citet{Tesfa2011} will be too large to fit into a single worker's memory, so
any approach based on this cannot scale, though it is generally possible to
convert a strip approach into a tiled approach~\citep{Yildirim2015}.

Frequent internode communication, as employed by many of the aforementioned
algorithms, is necessary to synchronize the nodes, but can slow down the
progression of the algorithm. More problematically, existing algorithms (except
for that of \citet{Yildirim2015}) require that enough nodes be available to hold
the entire DEM in RAM (otherwise a tile-swapping approach, prone to the
aforementioned problems, would be required). Since node count is a factor in
supercomputer scheduling, delays in the commencement of calculations may
dominate the time-to-solution. As an algorithm progresses towards a solution,
many nodes may be functionally idle, which unnecessarily wastes supercomputing
service units and monopolises resources.

The algorithm presented here is superior to existing parallel computing
approaches because it can (a)~guarantee that all nodes remain fully utilised
(save for a fixed number of brief synchronization events), (b)~it can operate
using fewer nodes than would be required to hold the entire dataset, and (c)~it
requires only a fixed number of internode communications and disk accesses. This
results in significant performance improvements over an existing algorithm.

\section{The Algorithm}
\label{sec:alg_overview}
Earlier, a depression-filled surface $W$ was defined. The effect of the
algorithm is to produce this surface, which will be referred to as the
\textit{global solution}. Since I am considering DEMs too large to fit into RAM
all at once, tiles will be used to calculate intermediate solutions which,
together, can be used to construct the global solution.

The algorithm has a single-producer, multiple-consumer design which proceeds in
three stages. (1)~The producer allocates tiles to the consumers, who calculate
an intermediate based on the tile and pass a small amount of information about
the intermediate back to the producer. (2)~Based on this data, the producer
calculates the information needed for each consumer to independently produce its
share of a global solution. (3)~It provides this to the consumers who modify
their intermediates based on it. The modified intermediates collectively form
the global solution: a depression-filled DEM. This design is effectively two
sequential MapReduce operations and is general enough to be implemented with
either threads or processes using any of a number of technologies including
OpenMP, MPI, Apache Spark~\citep{Zaharia2010}, or MapReduce~\citep{Dean2008}.
Here, I use MPI.

The third stage of the algorithm modifies intermediates generated by the first
stage. But this modification cannot take place until after the second stage has
completed. There are three strategies for caching these intermediates which
affect both the speed and the RAM requirements of the algorithm as a whole.
These strategies are as follows. (a)~The \textsc{evict} strategy: a consumer
evicts its intermediates from RAM and works on other tiles. This option uses the
least RAM and disk space. (b)~The \textsc{cache} strategy: a consumer writes its intermediates
to disk and works on other tiles. There is a related strategy, \textsc{cacheC}
in which the intermediate data is compressed before being written to disk. This strategy uses the same RAM as \textsc{evict}, but more disk space. Which strategy is fastest will depend on hardware configurations and
should be determined by testing. (c)~The \textsc{retain} strategy: a consumer
keeps its intermediate in RAM at all times.

If the DEM cannot fit entirely into the RAM of the available node(s), the
\textsc{evict} and \textsc{cache} strategies still allow the DEM to be
processed. In the limit, only the producer's information and a single tile need
be in RAM at a time. \textit{This allows large DEMs to be efficiently processed by a single-core machine}, decreasing resource costs and democratizing analysis. Additional RAM and cores, as may be available on high-end desktops or supercomputers, will result in faster time-to-completion. Only if sufficient RAM is available such that the entire dataset can be stored in RAM at once, can the \textsc{retain} strategy can be used. This strategy will result in the fastest time-to-completion.

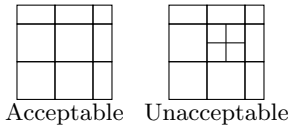
\begin{figure}
\centering
\begin{tikzpicture}[scale=0.5]
\draw (0,0) rectangle (1,1);
\draw (1,0) rectangle (2,1);
\draw (0,1) rectangle (1,2);
\draw (1,1) rectangle (2,2);
\draw (2,0) rectangle (2.5,1);
\draw (2,1) rectangle (2.5,2);
\draw (2,2) rectangle (2.5,2.5);
\draw (0,2) rectangle (1,2.5);
\draw (1,2) rectangle (2,2.5);
\node at (1.25,-0.4) {\small Acceptable};

\begin{scope}[shift={(4,0)}]
\draw (0,0) rectangle (1,1);
\draw (1,0) rectangle (2,1);
\draw (0,1) rectangle (1,2);
\draw (1,1) rectangle (2,2);
\draw (2,0) rectangle (2.5,1);
\draw (2,1) rectangle (2.5,2);
\draw (2,2) rectangle (2.5,2.5);
\draw (0,2) rectangle (1,2.5);
\draw (1,2) rectangle (2,2.5);
\draw (1.5,1) -- (1.5,2);
\draw (1,1.5) -- (2,1.5);
\node at (1.25,-0.4) {\small Unacceptable};
\end{scope}
\end{tikzpicture}
\caption{An example of an acceptable and an unacceptable tiling.\label{fig:tilings}}
\end{figure}

To proceed, the DEM is first subdivided into rectangular tiles. These need not
all have the same dimensions, but any two adjacent tiles must share the entire
length of their adjoining edges, as exemplified by Figure~\ref{fig:tilings}.
Relaxing these restrictions would be straight-forward, but is not done here in
order to maintain simplicity of presentation.

If the DEM comes in a pre-tiled form, as is the case with the datasets
considered here, these tiles can be used without modification as long as they
meet the aforementioned requirements. If the DEM is not pre-tiled, i.e.\ comes
in a single file, appropriate tile dimensions can be specified by the user and
portions of the file can then be read as tiles.

\begin{figure*}
\includegraphics[width=\textwidth]{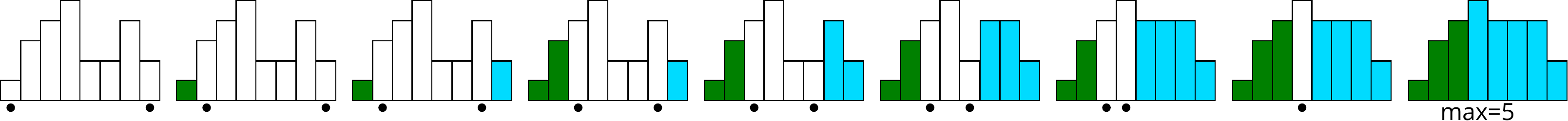}
\caption{Solving a single tile. The Priority-Flood begins by adding all of the
edge cells to the priority queue. Queued cells are represented by a black
circle. Each edge cell is the mouth of its own watershed, represented with
different colours here. The queue's lowest cell $c$ is dequeued and its
neighbours added to the queue; the neighbours inherit $c$'s watershed label.
Depressions are filled in. When two different watersheds meet, the maximum
elevation of the two meeting cells is noted: here there are five distinct
elevation levels and the two watersheds meet at an elevation of 5. If this
noted elevation is the lowest of any meeting of the two watersheds, it is
retained as the watersheds' spillover elevation.  Further details are
provided in~\citep{Barnes2014pf}. \label{fig:single_tile}}
\end{figure*}

\subsection{Solving a Single Tile}
\label{sec:single_block_solve}

In each tile, all depressions are filled and each cell is associated with a
``watershed": a collection of cells which all drain to the same outlet cell.
This operation is performed by the watershed variant of the \citet{Barnes2014pf}
Priority-Flood algorithm and applied to each tile, as described in the next
paragraph. \citet{Zhou2015} have recently published a new variant of Priority-
Flood which runs almost twice as fast as that presented by
\citeauthor{Barnes2014pf} The \citeauthor{Zhou2015} algorithm is too complex to
present here, but minor modifications make it a direct substitute for the
\citeauthor{Barnes2014pf} algorithm, so I use the former for timing tests and
the latter for description. If, in the future, even faster algorithms than that
of \citeauthor{Zhou2015} emerge, these could likewise be used. %which is similar to one described by \citet{Domingue1988},

The \citeauthor{Barnes2014pf} algorithm adds DEM cells to a priority-queue
\textsc{pq} which efficiently (in $O(\log N)$ time per cell) orders the cells
such that the cell with the lowest elevation is always at the front of
\textsc{pq}; \textsc{NoData} cells are taken to be lower in elevation than any
cell with data. The algorithm is initialized by adding all of the DEM's edge
cells to \textsc{pq}.

When a cell $c$ is popped from \textsc{pq}, if $c$ does not already have a
watershed label then its neighbours $n_i$ are considered. $c$ will be given the
label of the first $n_i$ (if any) found which already has a watershed label and
is at an elevation less than or equal to $c$. If no such $n_i$ is found, then
$c$ is given a new label.

Next, the elevation of any unlabeled $n_i$ which has an elevation less than $c$
is increased to match that of $c$. This step fills in depressions because
\textsc{pq} guarantees that any cell which is lower than $c$ and not part of a
depression would have been visited before $c$.

All $n_i$ which had not been previously labeled are given the same label as
$c$, indicating they are now part of the same watershed. For all of the $n_i$
which had been previously labeled, the maximum elevation of $c$ and $n_i$ is
noted and, if $n_i$ has a different label from $c$ and this elevation is less
than the elevation of any previously observed meeting of the two labels, it is
retained. This is the lowest spillover point from $c$'s watershed to $n$'s
watershed. Cumulatively, all of the spillover points form a \textit{spillover
graph} connecting watersheds together.

Finally, all $n_i$ which had not been previously labeled are added to
\textsc{pq} and the process repeats.

\begin{figure*}
\hfil%
\resizebox {0.3\textwidth} {!} {
\includegraphics[width=\textwidth]{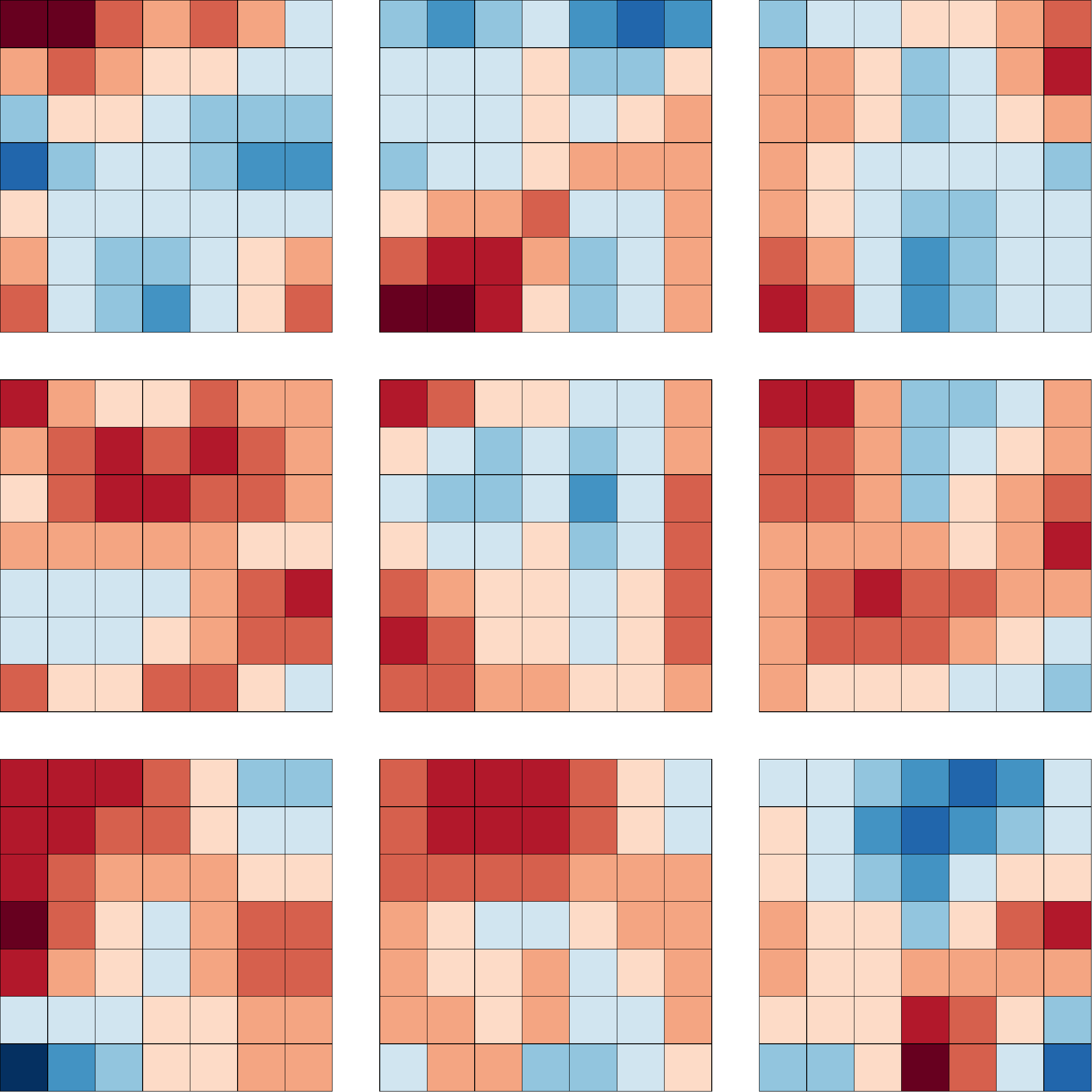}
}%
\hfil%
\resizebox {0.3\textwidth} {!} {
\includegraphics[width=\textwidth]{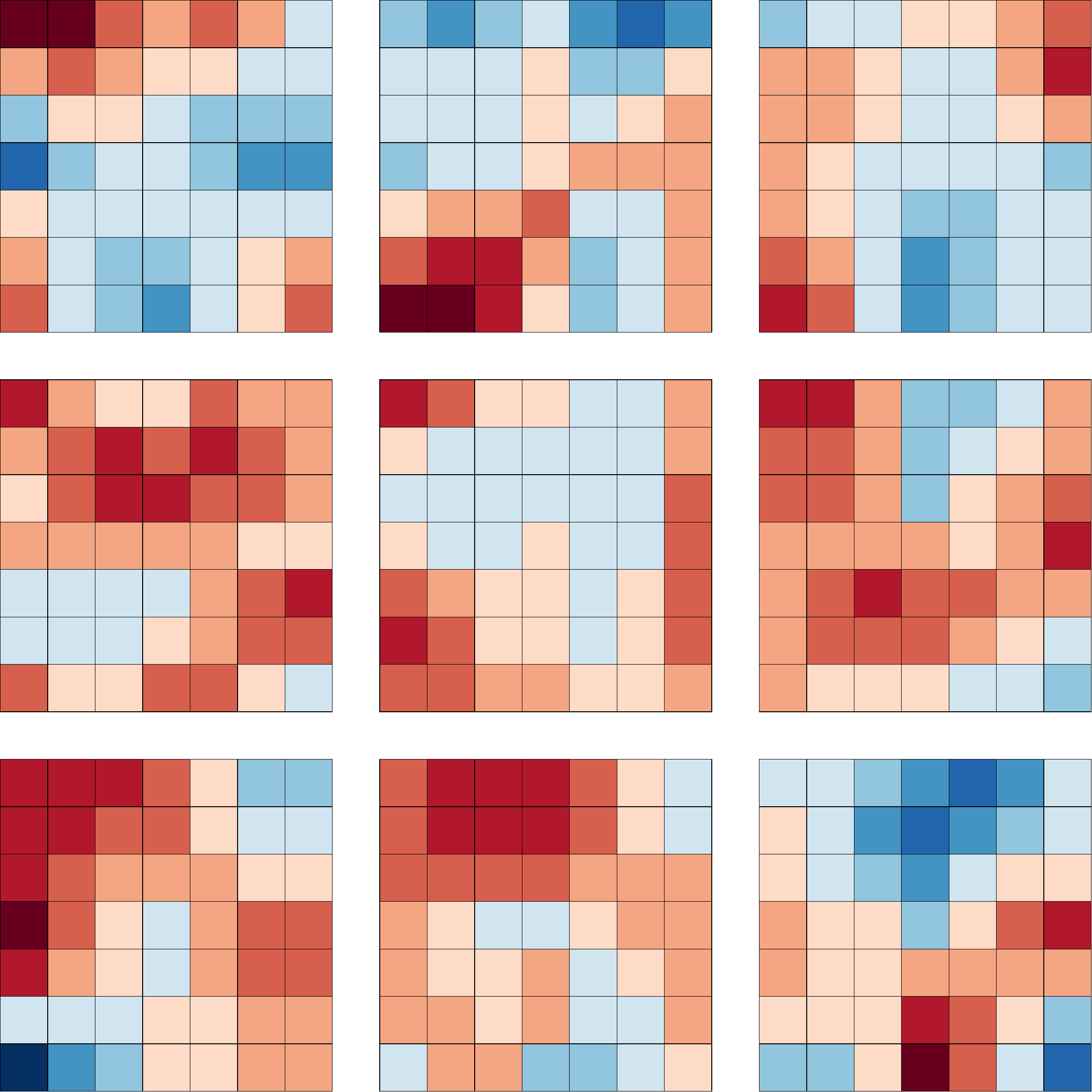}
}%
\hfil%
\resizebox {0.3\textwidth} {!} {
\includegraphics[width=\textwidth]{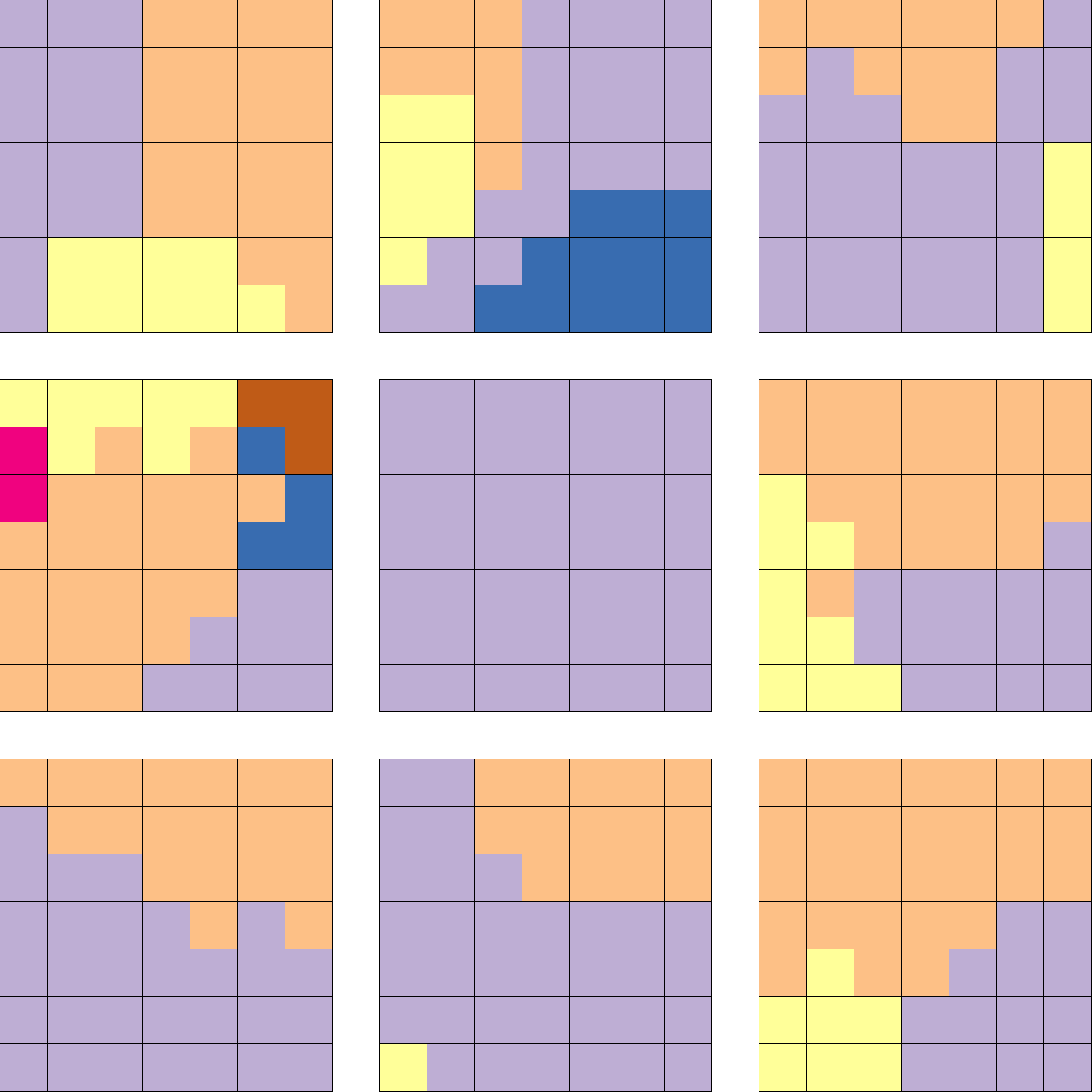}
}%
\hfil%

\vspace{-0.5em}
\hfil a \hfil b \hfil c \hfil

\vspace{0.5em}

\hfil%
\resizebox {!} {0.2in} {
\includegraphics[width=\textwidth]{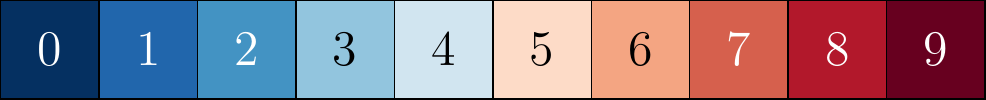}
}%
\hfil%
\caption{Global View of Solving a Single Tile. Cells are shown as small squares
with black borders and tiles as larger 7\,x\,7 squares separated by white space.
Colours in (a) and (b) correspond to elevations, as shown in the legend. Colours
in (c) correspond to various watershed labels; even though the same label colour
may appear in separate tiles each label should be considered globally unique.
(a) shows the raw DEM. In (b) the Priority-Flood depression filling operation
described in \textsection\ref{sec:single_block_solve} and
Figure~\ref{fig:single_tile} has been performed. The effect of this is that the
tiles no longer have internal depressions; this difference is most notable in
the central tile. Another effect of this is that each tile is now associated
with a ``watershed": a set of cells which drain to a common point. These
watersheds are shown in (c). Note that although the central tile does not
contain depressions, many of the cells in the central tile are part of a
depression when the DEM is considered as a whole.
\label{fig:step1}}
\end{figure*}

Once there are no more cells in \textsc{pq}, this step of the process is done.
At this point, if the tile being considered was adjacent to one of the edges of
the DEM, all of the cells' labels on that edge are noted as being connected via
their minimum elevation to the special label~1. This same information is
depicted graphically in Figure~\ref{fig:single_tile}, in pseudocode in
Algorithm~\ref{alg:subdiv_pf}, and with extensively commented supplementary
source code. The net effect of performing this step on each tile is shown in
Figure~\ref{fig:step1}.

\begin{breakablealgorithm}
\caption{\small {\sc Subdivision Priority-Flood}: This is a variation of
Algorithm 5 of \citet{Barnes2014pf}. A plain queue is used to accelerate the
standard Priority-Flood and a common label is applied to all cells draining to
an outlet.
\textbf{Upon entry,} \textbf{(1)}~\textit{DEM} contains the elevations of every
cell or the value \textsc{NoData} (which is assumed to be a very negative
number) for cells not part of the DEM. \textbf{(2)}~\textit{DEM} may be a tile
of a larger DEM. \textbf{At exit,} \textbf{(1)}~\textit{DEM} contains no
depressions. (2)~\textit{Labels} contains a label for every cell.
\textbf{(3)}~All cells which drain to a common point at the edge of the DEM bear
the same label. \textbf{(4)}~Graph associates label pairs with the minimum
spillover elevation between the labels.}
\label{alg:subdiv_pf}
\footnotesize
\begin{algorithmic}[1]
  \State Let \textit{Tile} be tile info from Algorithm~\ref{alg:main}.
  \State Let \textit{Open} be a min-first priority queue
  \State Let \textit{Pit} be a plain queue
  \State Let \textit{Labels} have the same dimensions as \textit{DEM}
  \State Let \textit{Labels} be initialized to 0
  \State Let \textit{Graph} associate label pairs with elevations
  \State
  \State Receive \textit{Tile} from the producer
  \State Read the specified portion of the full DEM into \textit{DEM}
  \ForAll{$c$ on the edges of \textit{DEM}}
    \State Push $c$ onto \textit{Open} with priority \textit{DEM}($c$)
  \EndFor
  \While{either \textit{Open} or \textit{Pit} is not empty}
    \If{\textit{Pit} is not empty}
      \State $c\gets\textsc{pop}(\textit{Pit})$
    \Else
      \State $c\gets\textsc{pop}(\textit{Open})$
    \EndIf

    \If{$\textit{Labels}(c)=0$}
      \If{$\exists$ a neighbour $n$ s.t.\ $\textit{Labels}(n)\ne0$ \textbf{and} $\textit{DEM}(n)\le\textit{DEM}(c)$}
        \State $\textit{Labels}(c)\gets\textit{Labels}(n)$
      \Else
        \State $\textit{Labels}(c)\gets\textsc{uniquelabel()}$
      \EndIf
    \EndIf

    \ForAll{neighbors $n$ of $c$}
      \If{$\textit{Labels}(n)\ne0$}
        \If{$\textit{Labels}(c)=\textit{Labels}(n)$}
          \State \Continue
        \EndIf
        \State $e\gets \max\big(\textit{DEM}(c),\textit{DEM}(n)\big)$
        \State $\textit{oe}\gets\textit{Graph}\big(\textit{Labels}(c),\textit{Labels}(n)\big)$
        \If{$oe=\textsc{null}$ \textbf{or} $e<oe$}
          \State $\textit{Graph}\big(\textit{Labels}(c),\textit{Labels}(n)\big)\gets e$
        \EndIf
      \Else
        \State $\textit{Labels}(n)\gets\textit{Labels}(c)$
        \If{$\textit{DEM}(n)\le c.z$}
          \State $\textit{DEM}(n)\gets c.z$
          \State Push $n$ onto \textit{Pit} with $z=c.z$
        \Else
          \State Push $n$ onto \textit{Open} with priority \textit{DEM}($n$)
        \EndIf
      \EndIf
    \EndFor
  \EndWhile
  \If{this was an edge tile}
    \ForAll{cells $c$ on the edge of the DEM}
      \State $\textit{oe}\gets\textit{Graph}\big(\textit{Labels}(c),1\big)$
      \If{$oe=\textsc{null}$ \textbf{or} $DEM(c)<oe$}
        \State $\textit{Graph}\big(\textit{Labels}(c),1\big)\gets DEM(c)$
      \EndIf
    \EndFor
  \EndIf
  \State Return edges of \textit{DEM} and \textit{Labels}, along with
  \textit{Graph} in a \textit{Tile} to the producer
\end{algorithmic}
\end{breakablealgorithm}

\begin{algorithm}
\footnotesize
\caption{\small {\sc HandleEdge}: Combine two tiles by joining their edges. An
analogous algorithm for handling corners is not shown.
\textbf{Upon entry,}
\textbf{(1)}~\textit{DEMA} and \textit{DEMB} contain the elevations of an
adjoining edge of the tiles A and B. \textit{LabelsA} and \textit{LabelsB}
contain the labels of an adjoining edge of the tiles A and B. \textit{Graph} is
a master graph containing the partially-joined graphs of all of the tiles. It
is modified in place. \textbf{At exit,}
\textbf{(1)}~\textit{DEMA}, \textit{DEMB}, \textit{LabelsA}, and
\textit{LabelsB} are unmodified. \textit{Graph} associates labels between the
two tiles with the minimum elevation required to spill between them.}
\label{alg:handle_edge}
\begin{algorithmic}[1]
  \State Let \textit{$DEMA$} be a vector of cell elevations from tile A
  \State Let \textit{$LabelsA$} be a vector of cell labels from tile A
  \State Let \textit{$DEMB$} be a vector of cell elevations from tile B
  \State Let \textit{$LabelsB$} be a vector of cell labels from tile B
  \State Let \textit{$Graph$} be an association of pairs of labels with an
  elevation
  \ForAll{$i$ in $\textsc{Length}(\textit{DEMA})$}
    \ForAll{$ni\in i+\{-1,0,1\}$}
      \LineIf{$ni<0$}{\Continue}
      \LineIf{$ni=\textsc{Length}(\textit{DEMA})$}{\Continue}
      \LineIf{$\textit{LabelsA}(i)=\textit{LabelsB}(ni)$}{\Continue}
      \State $e\gets \max\big(\textit{DEMA}(i),\textit{DEMB}(ni)\big)$
      \State $oe\gets \textit{Graph}\big(\textit{LabelsA}(i),\textit{LabelsB}(ni)\big)$
      \If{$oe=\textsc{null}$ \textbf{or} $e<oe$}
        \State $\textit{Graph}\big(\textit{LabelsA}(i),\textit{LabelsB}(ni)\big)\gets e$
      \EndIf
    \EndFor
  \EndFor
\end{algorithmic}
\end{algorithm}

\begin{figure}
\centering
\includegraphics[width=\columnwidth,height=1in,keepaspectratio]{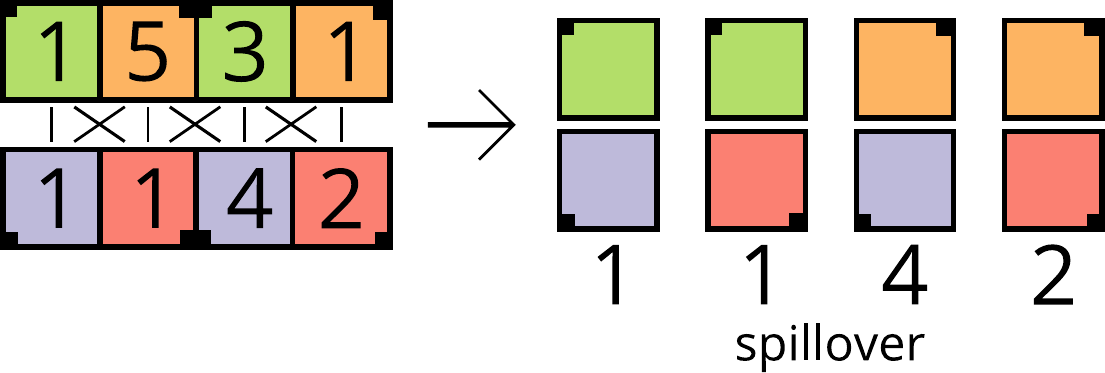}
\caption{Handle Edges. The adjoining edge cells of tiles are used to construct
a global solution. Here cells have elevations corresponding to the number in
their center and belong to watersheds denoted both by their colour and the
orientation of the small black box. All cells are compared with their neighbours
in the adjacent tile, as denoted by the black lines in the upper half of the
figure. After the algorithm is finished, the spillover elevation of all adjacent
watersheds is known, as depicted by the right-hand side of the figure. In each
cell-cell comparison, the maximum elevation of the pair is that pair's
spillover. For each watershed-watershed pair, the minimum of the cell-cell
comparisons is the watershed spillover. As an example, consider just one pair of
watersheds: $\min(\max(4,5),\max(4,1))=\min(5,4)=4$.
\label{fig:edge_handle}}
\end{figure}

\subsection{Constructing a Global Solution}
\label{sec:global_solution}

\begin{figure*}
\hfil%
\resizebox {0.3\textwidth} {!} {
\includegraphics[width=\textwidth]{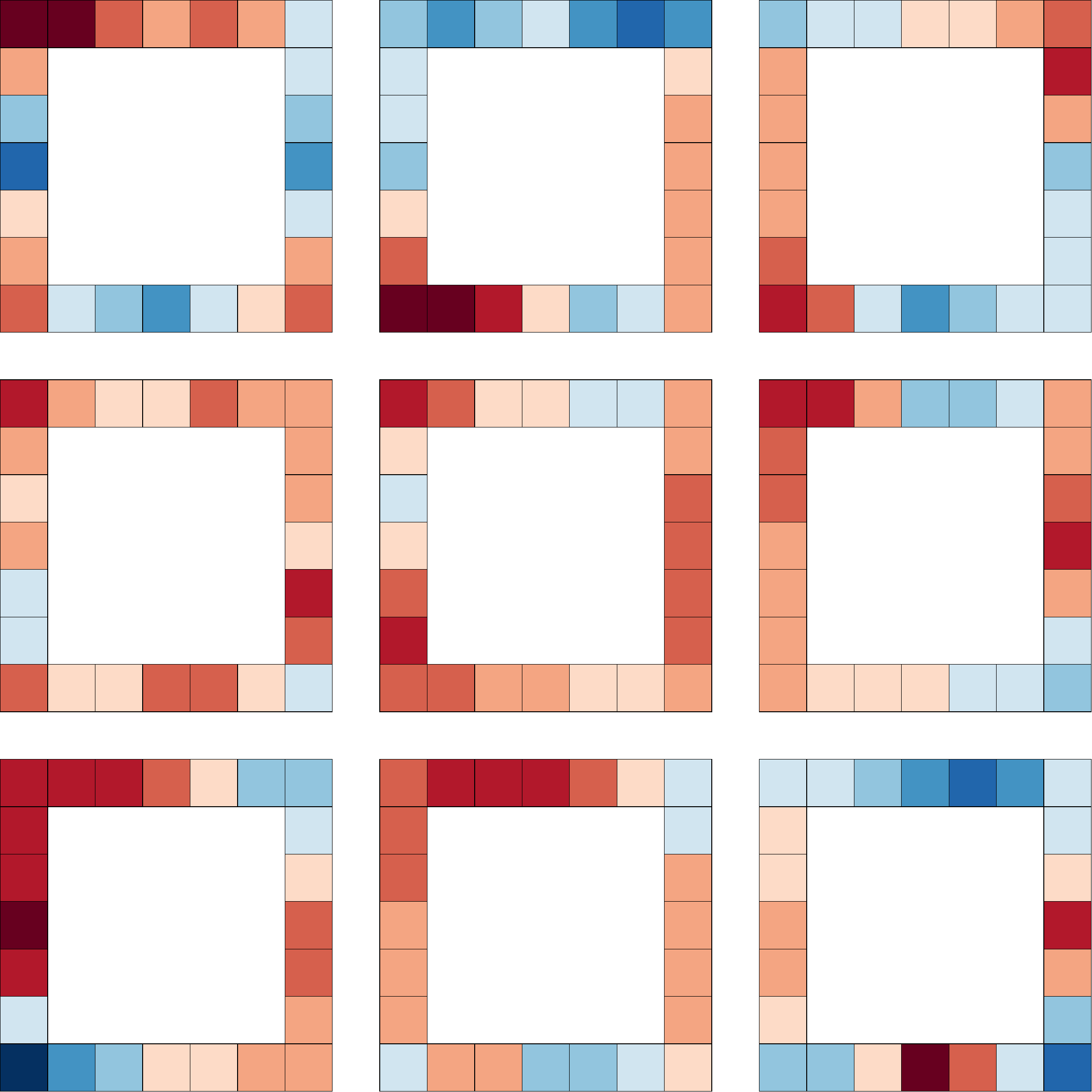}
}%
\hfil%
\resizebox {0.3\textwidth} {!} {
\includegraphics[width=\textwidth]{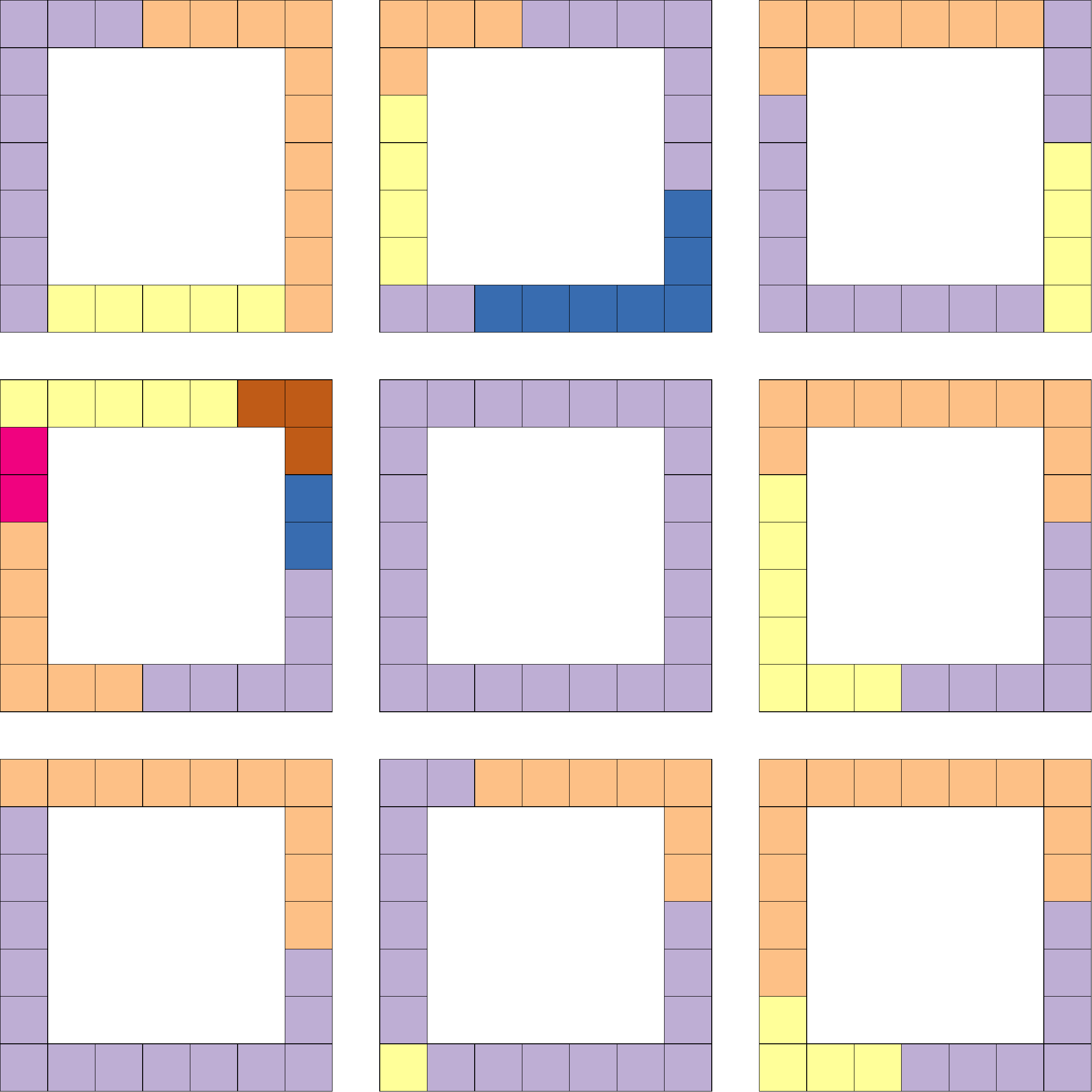}
}%
\hfil%
\resizebox {0.3\textwidth} {!} {
  \begin{tikzpicture}
    \node at (0in,0in) {\includegraphics[width=1in,height=1in,keepaspectratio]{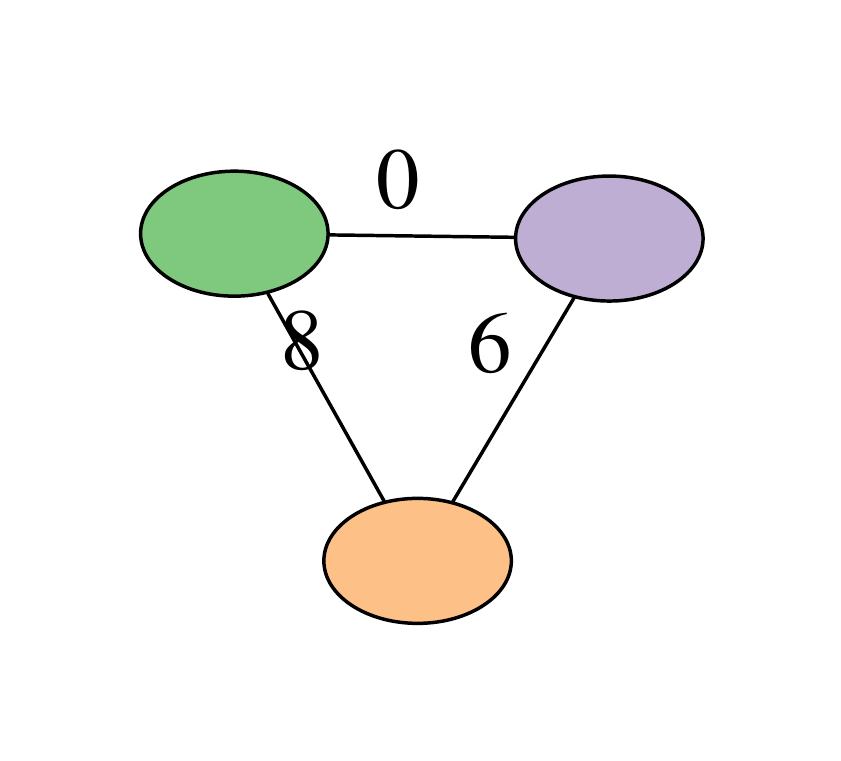}};
    \node at (1in,0in) {\includegraphics[width=1in,height=1in,keepaspectratio]{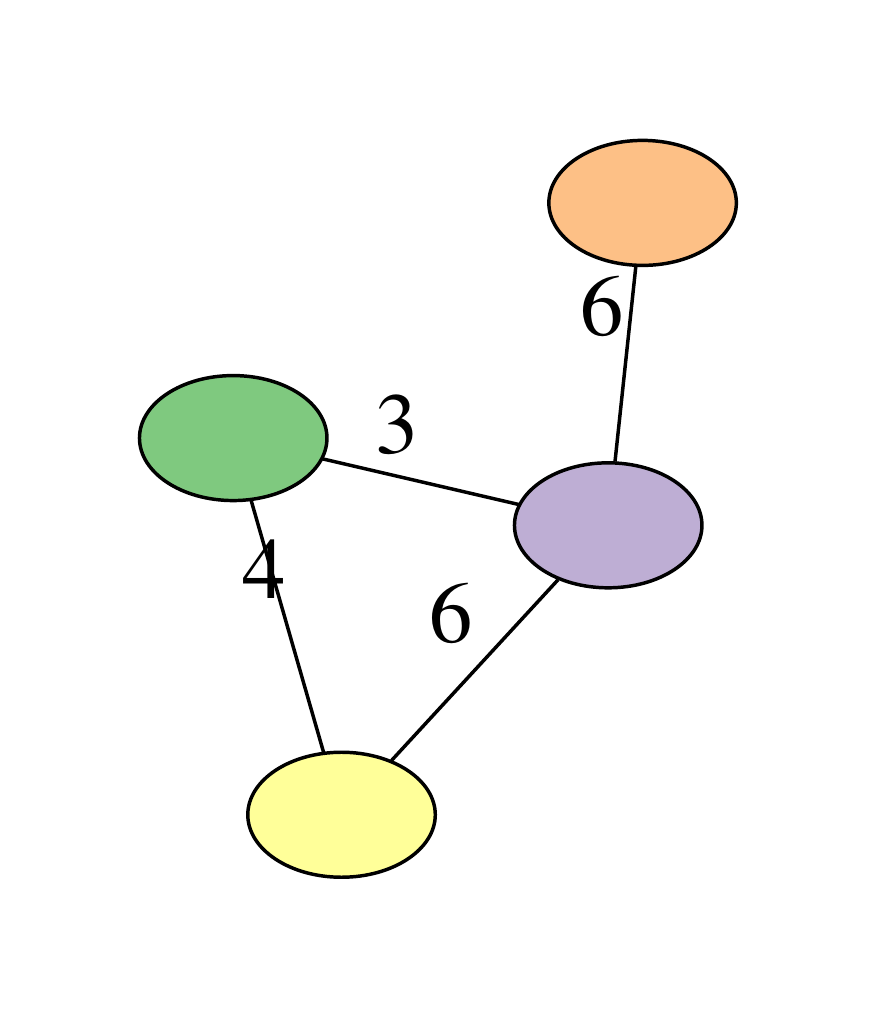}};
    \node at (2in,0in) {\includegraphics[width=1in,height=1in,keepaspectratio]{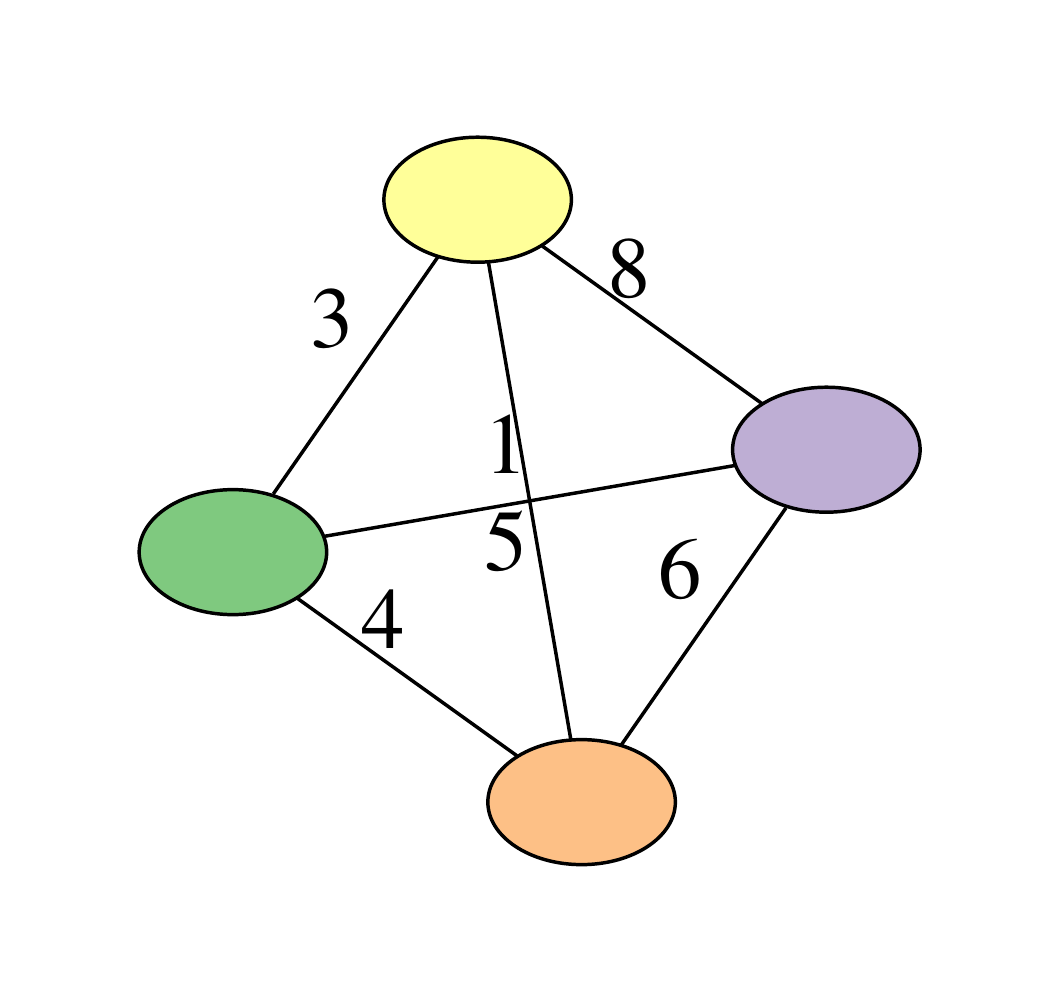}};
    \node at (0in,1in) {\includegraphics[width=1in,height=1in,keepaspectratio]{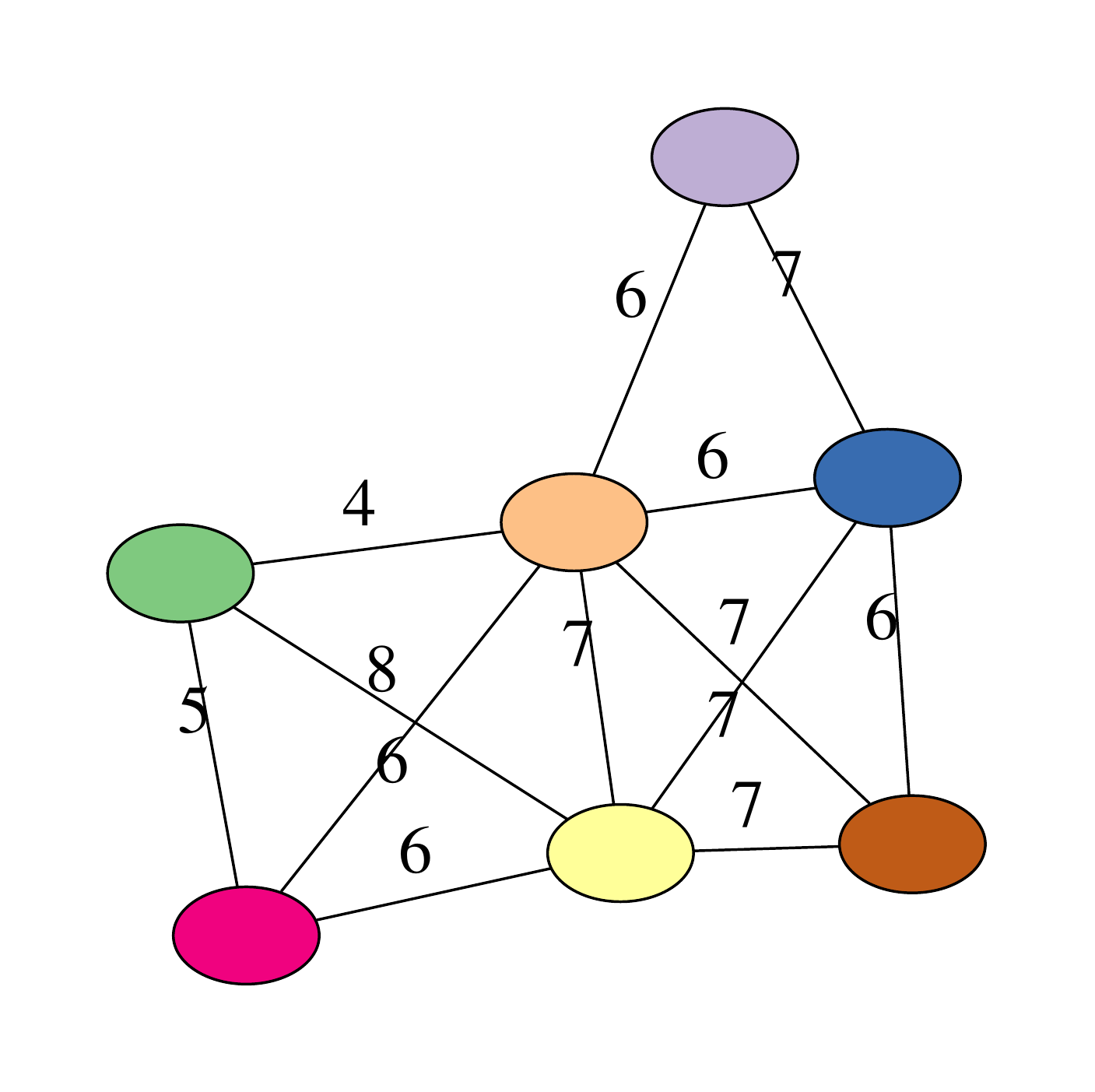}};
    \node at (1in,1in) {\includegraphics[width=1in,height=1in,keepaspectratio]{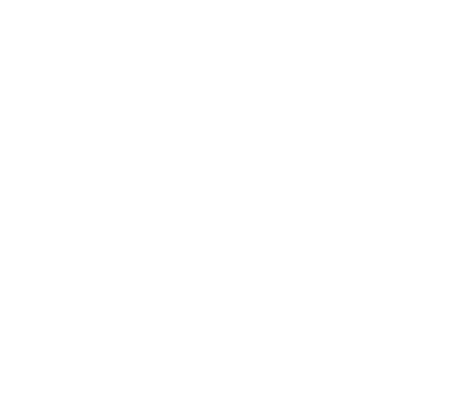}};
    \node at (2in,1in) {\includegraphics[width=1in,height=1in,keepaspectratio]{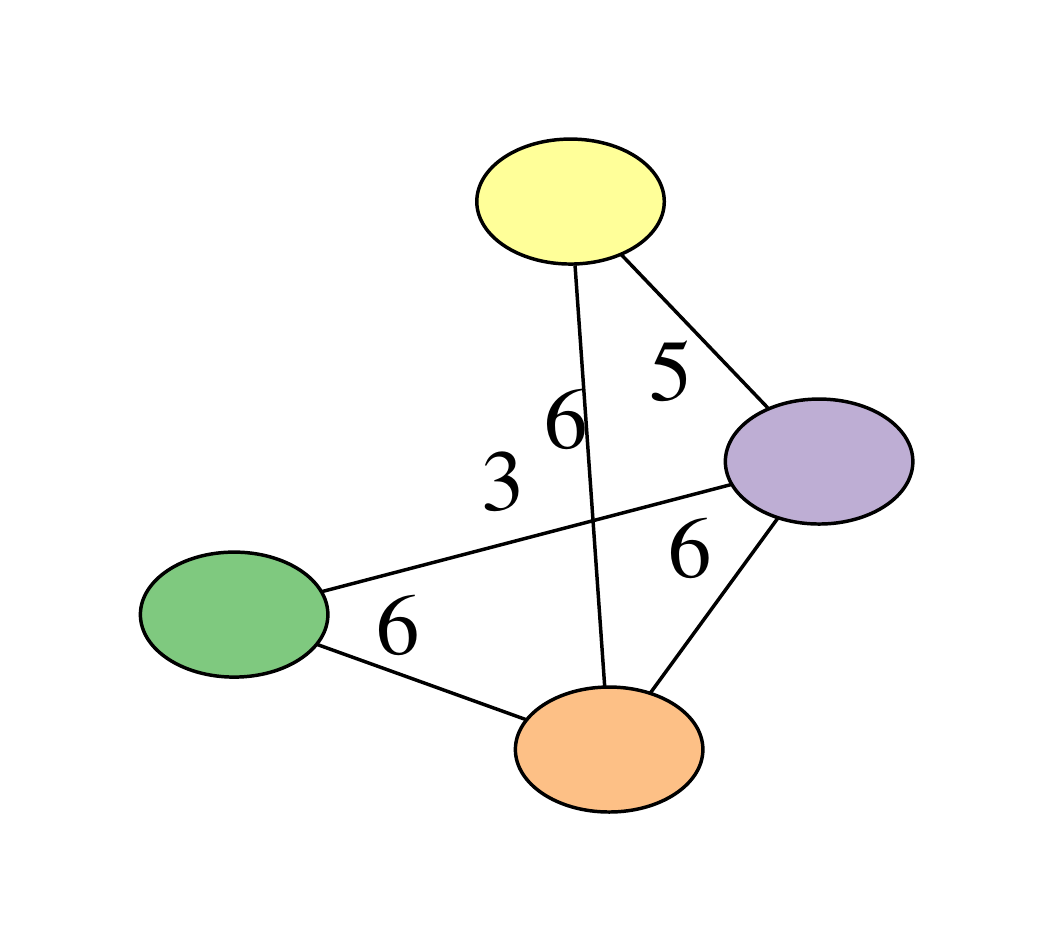}};
    \node at (0in,2in) {\includegraphics[width=1in,height=1in,keepaspectratio]{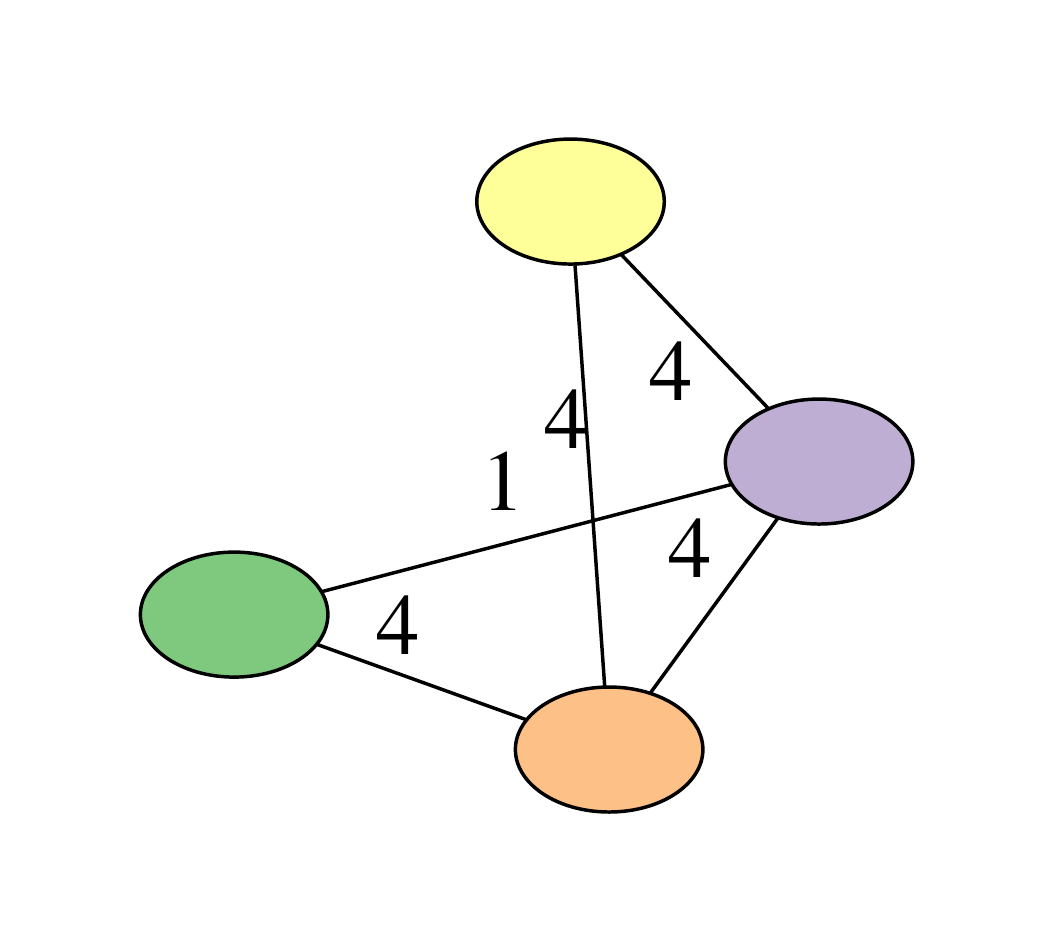}};
    \node at (1in,2in) {\includegraphics[width=1in,height=1in,keepaspectratio]{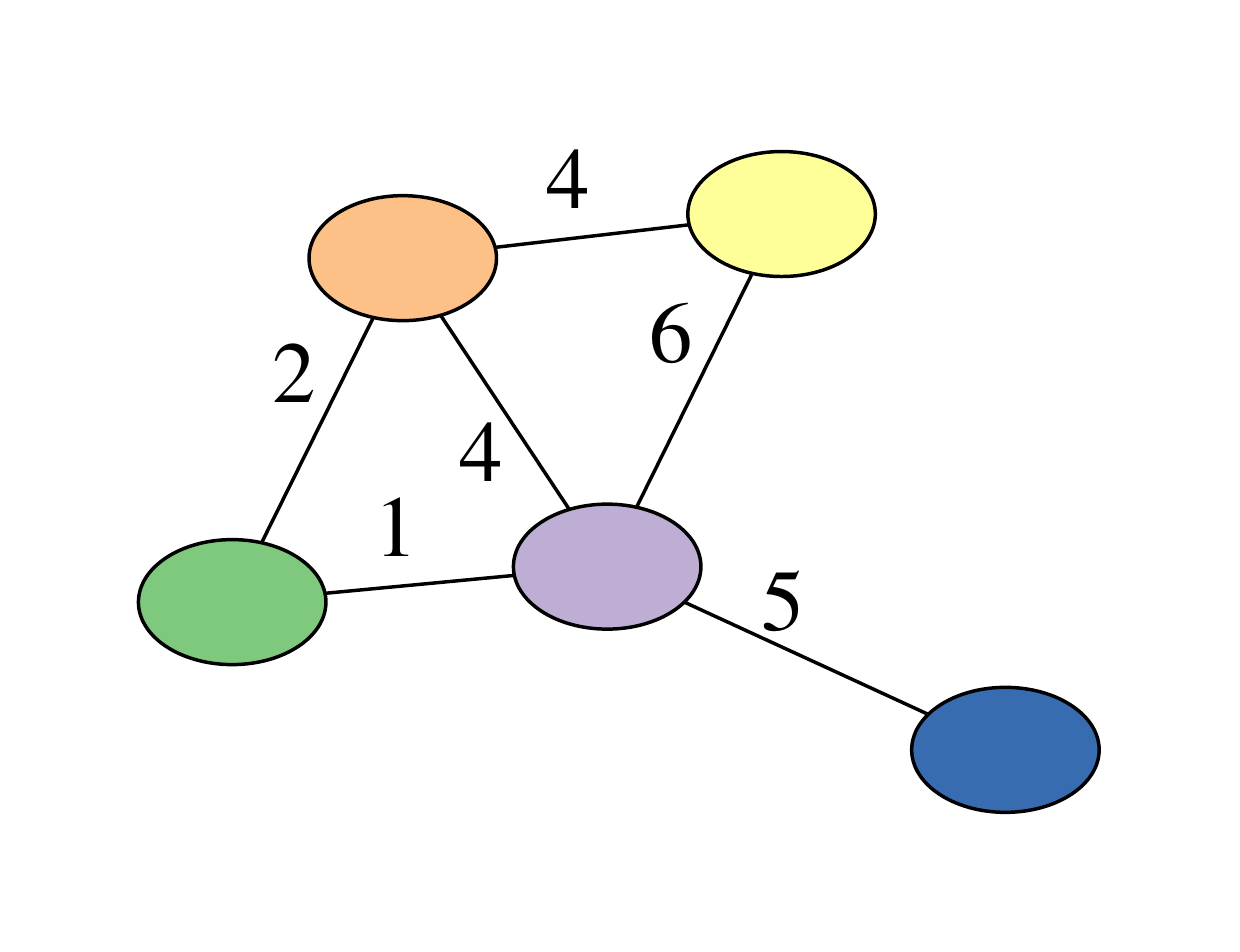}};
    \node at (2in,2in) {\includegraphics[width=1in,height=1in,keepaspectratio]{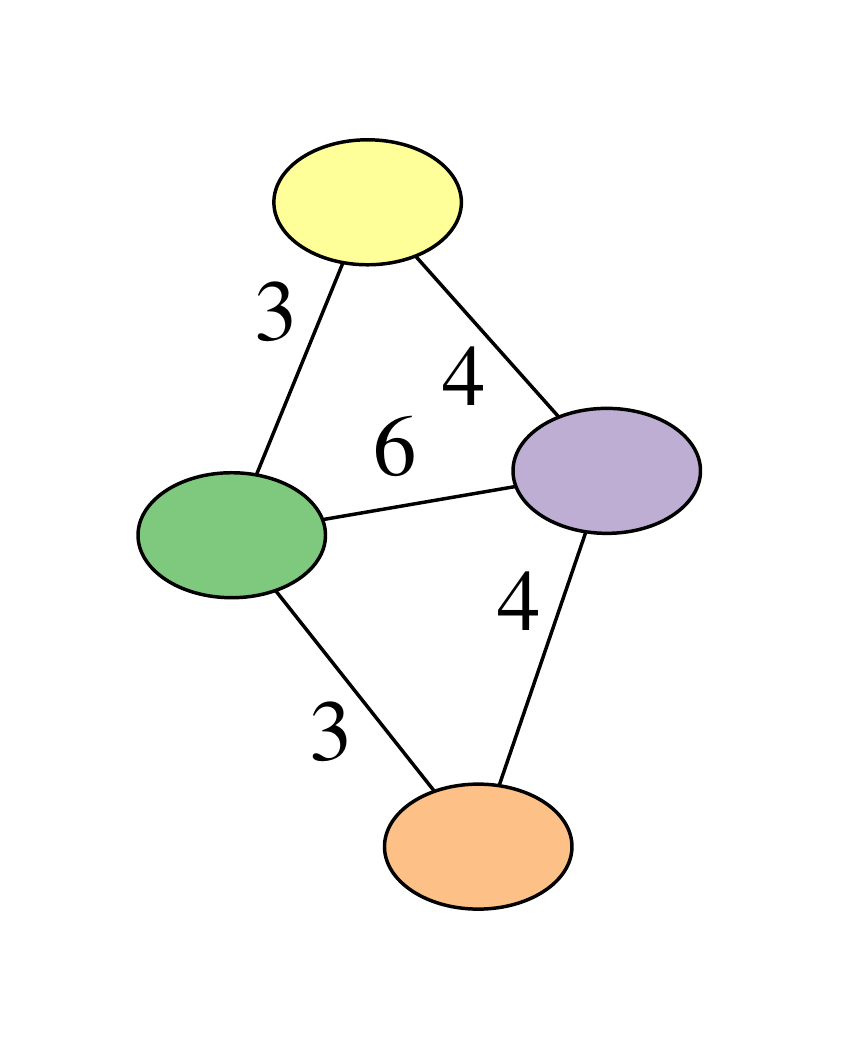}};
  \end{tikzpicture}
}%
\hfil%

\vspace{-0.5em}
\hfil a \hfil b \hfil c \hfil

\vspace{0.5em}

\caption{Communication Required for a Global Solution. (Refer to Figure~\ref{fig:step1} for an
explanation of colours.) A portion of the information shown in
Figure~\ref{fig:step1} is needed to construct a global solution. The elevations
of each perimeter cell (shown in (a)), the labels of each perimeter cell (shown
in (b)), and the spillover graphs of each tile (shown in (c)) are sent to a
central node. The central tile does not have a spillover graph because all of
the cells are part of the same watershed: a node for this watershed is created
later (see Figure~\ref{fig:masterstep}).
\label{fig:step1_comm}}
\end{figure*}

As each tile finishes being processed, as described above, its consumer sends
some information about the tile to the producer, as described in the next
paragraph. Once this information is sent, the consumer can apply one of the
caching strategies described above: \textsc{evict}, \textsc{cache}, or
\textsc{retain}. If \textsc{cache} or \textsc{retain} are used, the depression-
filled DEM and the watershed labels of each cell must be saved. The spillover
graph can be discarded.

The consumer sends the following information to the producer: (a)~the elevations
of each cell on all four edges of the tile, (b)~the labels of each cell on all
four edges of the tile, and (c)~the tile's spillover graph.
Figure~\ref{fig:step1_comm} depicts this. The amount of information sent is
therefore proportional to the length of the tile's perimeter and its number of
watersheds; all of this information is sent only once per tile. Communication
costs and data sizes are discussed theoretically in
\textsection\ref{sec:complexity} and empirically in
\textsection\ref{sec:results} and Table~\ref{tbl:results}.

The producer uses non-blocking communication to delegate unprocessed tiles to
consumers in round-robin fashion. The producer then uses a blocking receive to collect data from the consumers as they finish processing. Once all of the
tiles have been processed, the elevation and label information is used to merge
all of the tiles' spillover graphs into a single, large spillover graph
encompassing all of the watersheds in the DEM. To do this, the labels of each
tile are adjusted so that they are globally unique (except for the special label 1, which indicates the edge of the DEM) and each of the separate graphs are unioned into a large graph.

Next, each pair of adjoining edges is considered and used to connect the
individual tiles' spillover graphs together. Each cell $c$ of an edge is
adjacent to 2--3 neighbouring cells $n_i$ in its adjoining edge. For each pair
of cells $\{c,n_i\}$, the maximum elevation of the two cells is noted and
retained if the labels of the two cells differ and no previous meeting of the
two labels has generated a lower elevation. A similar procedure is performed
for the corner cells of tiles which are diagonally adjacent. This same
information is depicted graphically in Figure~\ref{fig:edge_handle}, in
pseudocode in Algorithm~\ref{alg:handle_edge}, and via extensive comments in
the supplementary source code.

\begin{figure*}
\hfil%
\includegraphics[width=0.3\textwidth,height=2.5in,keepaspectratio]{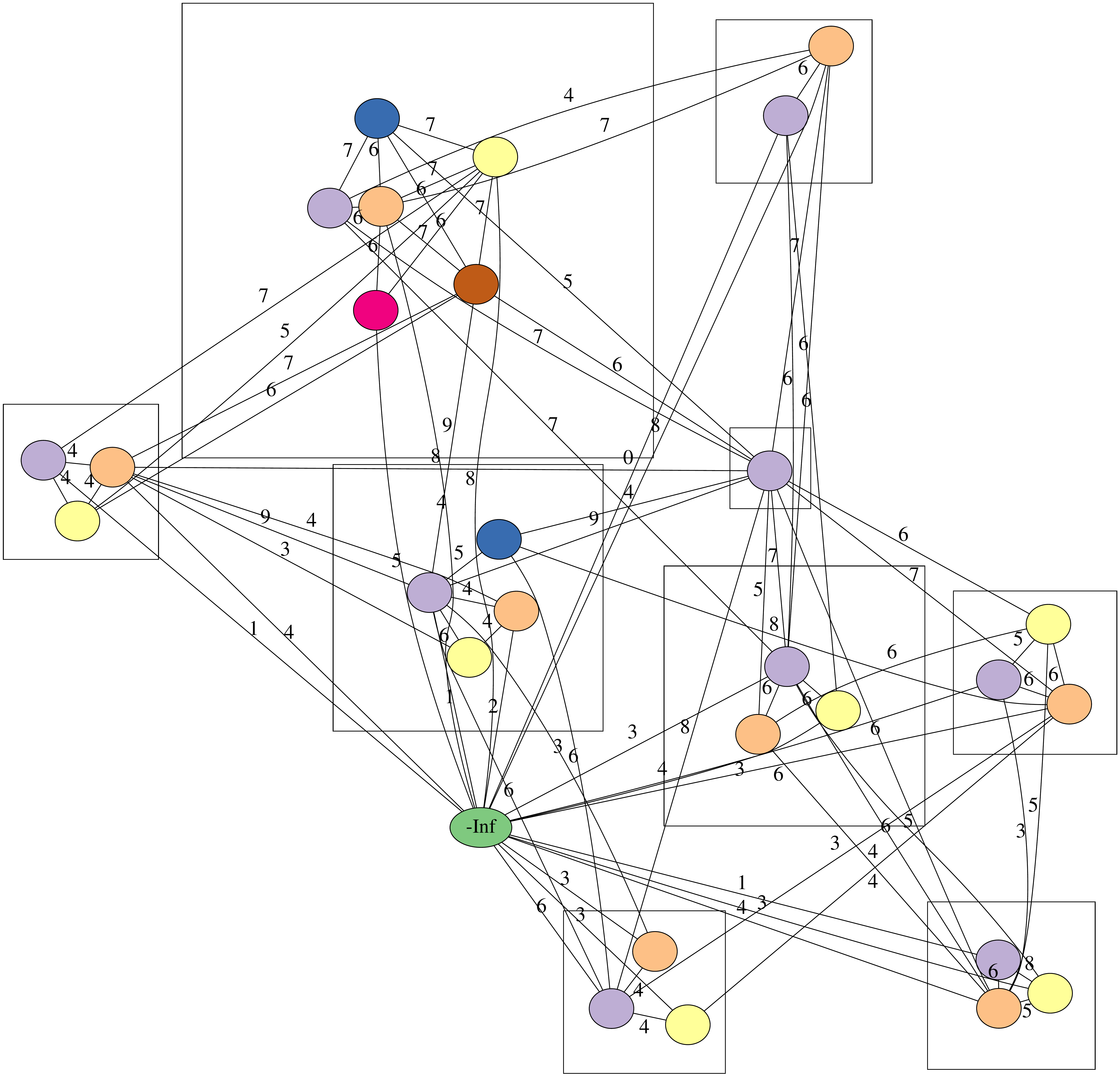}%
\hfil%
\includegraphics[width=0.3\textwidth,height=2.5in,keepaspectratio]{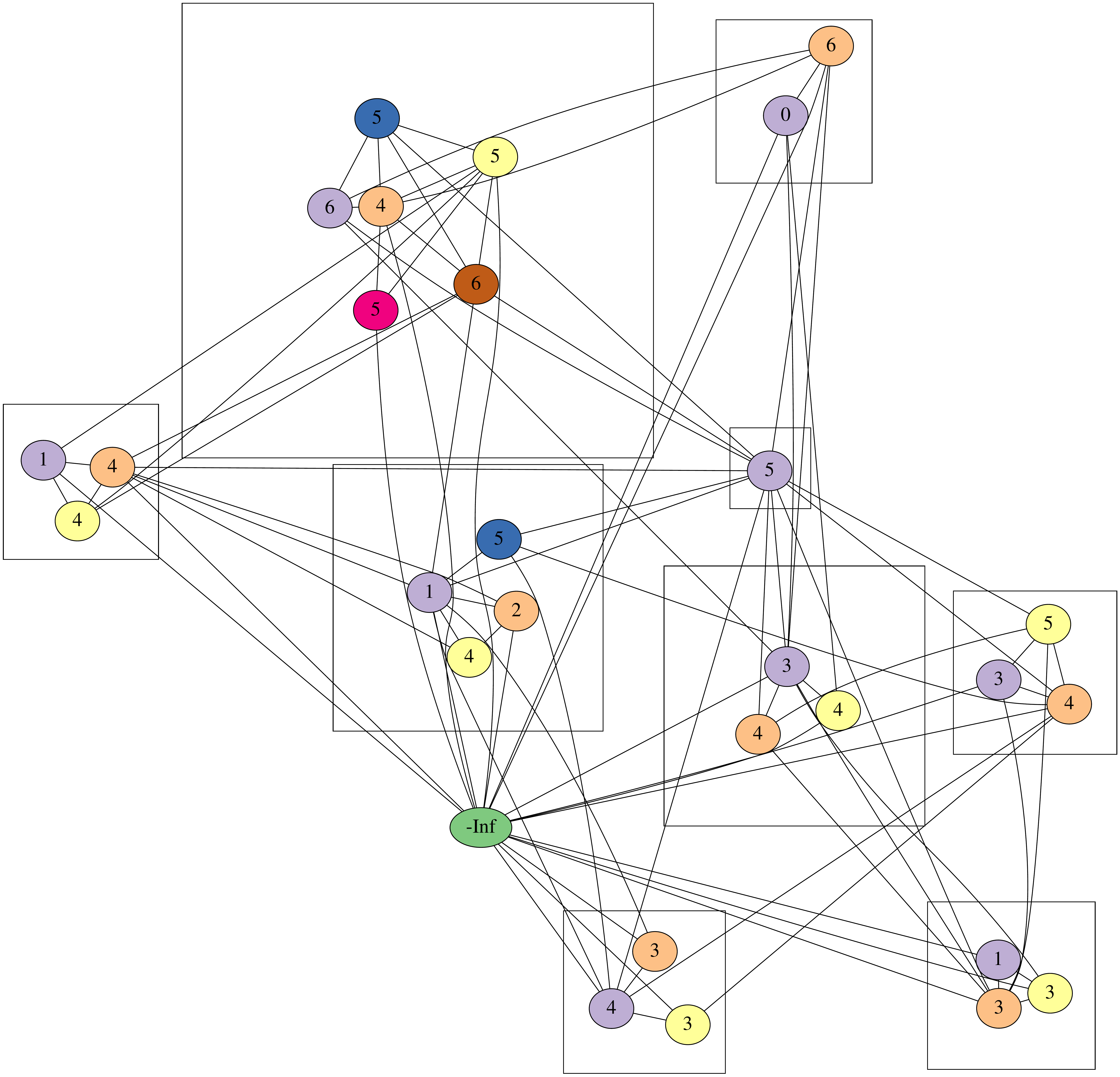}%
\hfil%

\hfil a \hfil b \hfil

\caption{Depression-Filling on the Global Spillover Graph. As described by
\textsection\ref{sec:global_solution}, the information communicated from each
tile, as shown in Figure~\ref{fig:step1_comm} is combined to form a global
spillover graph (shown in (a)). Priority-Flood depression-filling is performed
on this graph to determine the minimum elevations of each label (shown in (b)).
These graphs are too large to comfortably put in print, but can be read in
electronic versions of this paper. Figure~\ref{fig:single_tile} provides a
visual representation of how the Priority-Flood works. \label{fig:masterstep}}
\end{figure*}

The resulting large graph is itself a digital elevation model. All of the
watersheds (represented by nodes in the large graph) adjacent to the edges of
the DEM have been linked to a single node with the special label~1
(\textsection\ref{sec:single_block_solve}) which is taken to have an elevation
of $-\infty$. This is used to seed a Priority-Flood (Algorithm 2 from
\citet{Barnes2014pf}) which sets the elevation of each node of the spillover
graph to the level of the lowest spillover point by which that node can be
accessed. Figure~\ref{fig:masterstep} depicts this.

\begin{algorithm}
\footnotesize
\caption{\small{ {\sc Main Algorithm}: 
\textbf{Upon entry,}
\textbf{(1)}~\textit{DEM} contains the elevations of every cell or the value
\textsc{NoData} (a very negative number) for cells not part of the DEM.
\textbf{At exit,}
\textbf{(1)}~\textit{DEM} contains no depressions. Communication is assumed to be non-blocking, except where otherwise noted. Consumers perform their calculations asynchronously with respect to the Producer. Note that consumers must be assigned the same tiles in the first and second part of the algorithm for \textsc{retain} to work.}}
\label{alg:main}
\begin{algorithmic}[1]
  \State Let \textit{Consumers} be a thread/process pool
  \State Let a tile have the filename, dimensions, edge information, and
  spillover graph for a tile
  \State Let \textit{Tiles} be a collection of tiles
  \State Let \textit{MGraph} be a graph which associates pairs of labels with elevations
  \State Let \textit{DEM} be a rather large digital elevation model
  \State
  \State Divide \textit{DEM} into tiles
  \ForAll{tiles $b$}
    \State Delegate $b$ to the next consumer $t$
    \State Have $t$ perform Algorithm~\ref{alg:subdiv_pf} on $b$
    \State If there are no more consumers start again at the first
  \EndFor
  \While{any tile is still unreceived}
    \State Block until any consumer returns
    \State Store the information returned
  \EndWhile
  \State
  \State Make the labels of \textit{Tiles} globally unique
  \State Merge all graphs in \textit{Tiles} into \textit{MGraph}
  \State
  \ForAll{adjoining edges $e$ of adjacent tiles}
    \State Pass $e$ and \textit{MGraph} to Algorithm~\ref{alg:handle_edge}
  \EndFor
  \ForAll{adjoining corners $c$ of diagonally adjacent tiles}
    \State Pass $c$ and \textit{MGraph} to an Algorithm~\ref{alg:handle_edge} analogue
  \EndFor
  \State
  \State Run Alg.\ 2 from \citet{Barnes2014pf} on \textit{MGraph}'s labels
  \State Let \textit{MResult} be the elevation of each label after this
  \State Adjust \textit{MResult} back to tile-specific labels
  \State
  \ForAll{tiles $b$}
    \State Send $b$ and its portion of \textit{MResult} to the next consumer $t$
    \If{$t$ cached the results of Algorithm~\ref{alg:subdiv_pf}}
      \State Let $t$ load the cached results
    \Else
      \State Let $t$ rerun Algorithm~\ref{alg:subdiv_pf}
    \EndIf
    \State Let $t$ raise the elevation of cells to match \textit{MResult}
    \State If there are no more consumers start again at the first
  \EndFor
\end{algorithmic}
\end{algorithm}

\begin{comment}
\begin{figure*}
\centering
\begin{tikzpicture}
  \node at (0,0) {\includegraphics[width=2in]{grid_fig/elev.pdf}};
  \node at (2.5in,0) {\includegraphics[width=2in]{grid_fig/labels.pdf}};
  \node at (2.5in,-2.5in) {\includegraphics[width=2in]{grid_fig/elev_filled.pdf}};
\end{tikzpicture}
\caption{TODO \label{fig:grid_fig}}
\end{figure*}
\end{comment}

\subsection{Broadcasting \& Finalizing the Global Solution}
\label{sec:broadcasting}

\begin{figure*}
\hfil%
\resizebox {0.3\textwidth} {!} {
\includegraphics[width=\textwidth]{terrain0.pdf}
}%
\hfil%
\resizebox {0.3\textwidth} {!} {
\includegraphics[width=\textwidth]{terrain0_filled.pdf}
}%
\hfil%
\resizebox {0.3\textwidth} {!} {
\includegraphics[width=\textwidth]{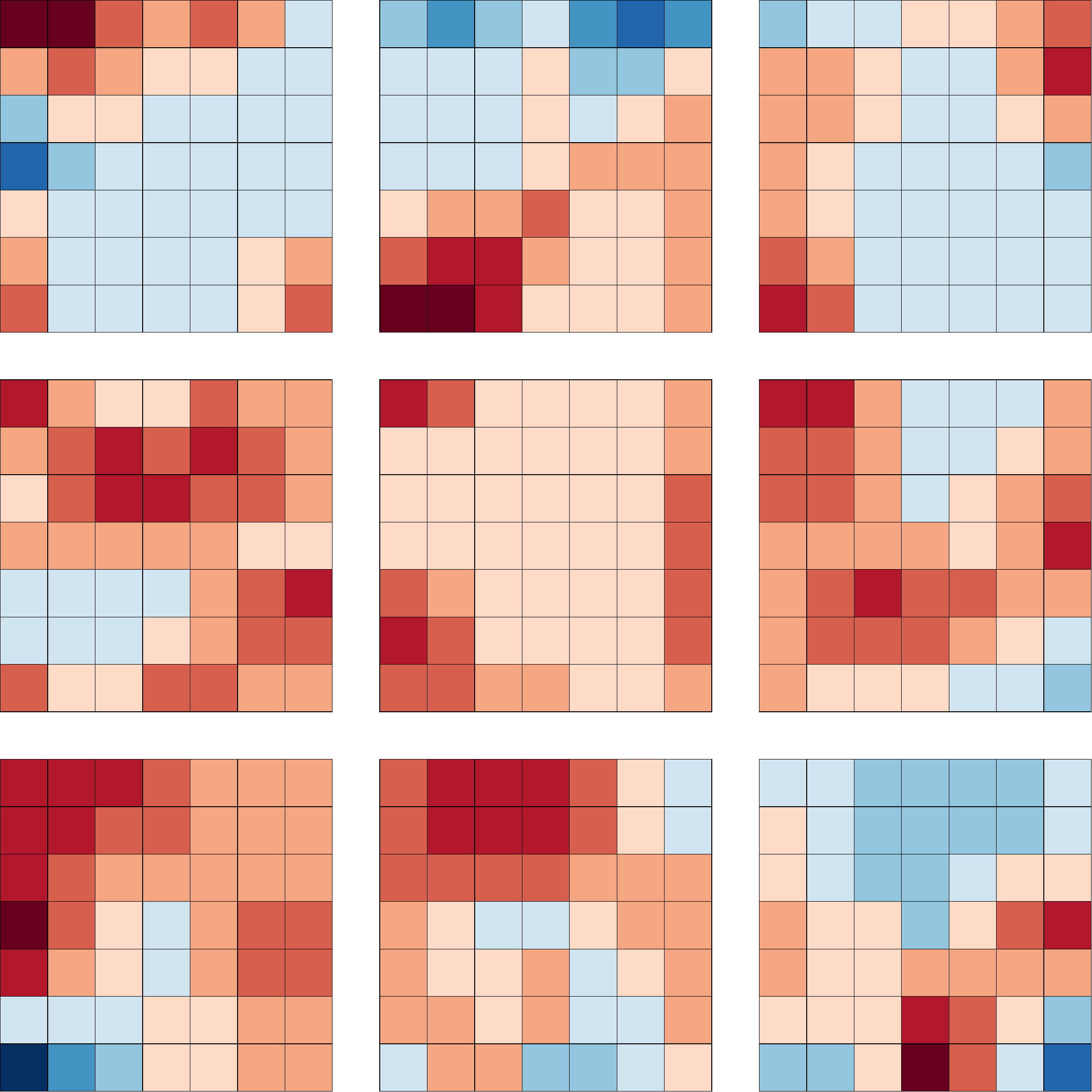}
}%
\hfil%

\vspace{-0.5em}
\hfil a \hfil b \hfil c \hfil

\vspace{0.5em}

\hfil%
\resizebox {!} {0.2in} {
\includegraphics[width=\textwidth]{scale.pdf}
}%
\hfil%
\caption{Progression to a Global Solution. (a) shows the raw DEM and (b) shows
the result of performing depression-filling on each individual tile, as in
Figure~\ref{fig:step1}. (c) shows the result of raising the each cell to the
minimum elevation of its label as determined by the depression-filled global
spillover graph shown in Figure~\ref{fig:masterstep}. (c) therefore represents
the desired, depression-filled DEM and contains no depressions at any scale.
Each tile can be saved separately or combined into a single output. The central
tile most clearly shows the progression.
\label{fig:step2}}
\end{figure*}

Recall that nodes represent watersheds, which may consist of cells of many
different elevations. The foregoing has established the minimum elevation any
cell in the watershed may have while still being guaranteed of draining to the
edge of the DEM. Therefore: any cell with that watershed's label below this
level must have its elevation increased. To accomplish this, the labels are
adjusted to once again be tile-specific. The labels and their associated global
elevations are then distributed to their respective tiles.

In order to perform this final elevation adjustment, each tile needs the
depression-filled elevations and labels generated in
\textsection\ref{sec:single_block_solve} by Algorithm~\ref{alg:subdiv_pf}. How
these are now obtained depends on the chosen caching strategy. If,
(a)~\textsc{evict} was used, then the intermediate must be recalculated as
described by \textsection\ref{sec:single_block_solve}, and then the
aforementioned adjustment made. Alternatively, if (b)~\textsc{cache} or
(c)~\textsc{retain} were used, then a single $O(N)$ scan of the tile is
sufficient to finalize the solution. The pros and cons of these strategies are
discussed in \textsection\ref{sec:complexity}.

Ultimately, each tile is saved separately to disk for further processing, which
may include mosaicing the tiles back into a single depression-filled DEM. The
foregoing information is encapsulated in Algorithm~\ref{alg:main} and via
extensive comments in the supplementary source code.

\section{Theoretical Analysis}
\label{sec:complexity}

\subsection{Time Complexity}
The time complexity of the algorithm is a function of the time taken to process
each individual tile and the time taken to build the global solution. Individual
tiles are processed using some variant of Priority-Flood. If $n$ is the number
of cells per tile, this takes $O(n+m \log m)$ time per tile ($O(n)$ for integer
data) where $m\le n$; typically, $m\ll n$. Let us assume the worst-case, which
implies $O(n\log n)$ time per tile.

The global solution requires that Priority-Flood be performed on the combined
spillover graph of the tiles. The number of nodes in this graph is proportional
to the number of watersheds. The maximum number of watersheds a tile can have is
equal to its number of edge cells, which is $\sim4\sqrt{n}$. If we call the
number of tiles $T$, then the global solution takes $O(T\sqrt{n} \log
T\sqrt{n})$.

Once this graph has been processed, if the individual tiles were cached, then an
$O(n)$ sweep per tile is sufficient to finish the job, otherwise, if
\textsc{evict} was used, Priority-Flood must be performed on each tile followed
by an $O(n)$ sweep. Assuming the worst case, finalizing takes $O(n \log n)$
time.

Therefore, in the worst-case, the total time is $O(Tn \log n)$ or $O(Tn)$ for
integer data. Either way, for a fixed tile size, the algorithm is linear in the
number of cells. Running a single Priority-Flood on the entire dataset at once
would take $O(Tn \log Tn)$ time ($O(Tn)$ for integer data)~\citep{Barnes2014pf};
therefore, the new algorithm should be faster even without employing multiple
cores. The aforementioned \citet{Planchon2002} algorithm operates in
$O((Tn)^{1.2})$ time; this is significantly slower than the new algorithm.

\subsection{Disk Access}

The new algorithm guarantees that each tile, and therefore, each cell, need only
be loaded into memory a fixed number of times. Recall from
\textsection\ref{sec:alg_overview} that there are three memory retention
strategies. (a)~\textsc{retain}. The entire dataset is retained in the memory of
the nodes at all times: this requires one read and one write per cell.
(b)~\textsc{cache}. The dataset cannot fit entirely into the memory of the
nodes, so intermediate results (labels and elevations) are cached to disk: this
requires three reads and three writes per cell. The \textsc{cacheC} strategy
would require less, but this is difficult to analyze due to the many compression
algorithms that could be used. (c)~\textsc{evict}. No intermediates are cached:
this requires two reads and one write per cell.

\textsc{retain} is the fastest strategy, but unlikely to be feasible for
large datasets. \textsc{cache} reduces computation versus \textsc{evict}, but is
more expensive in terms of disk access. \textsc{cacheC} may use nearly any
amount of computation depending on the algorithm employed: a good algorithm
should yield acceptable compression with minimal processing. Previous algorithms
based on virtual tiles must be at least as expensive as \textsc{retain}. Each
time such an algorithm swaps a virtual tile out of memory, it incurs the cost of
one write (and, later), one read. Therefore, if approximately half the virtual
tiles are swapped once, the costs will surpass \textsc{evict}. Put another way:
if the dataset is twice as large as the available RAM, it is reasonable to
expect a virtual tile algorithm to be more expensive than that presented here.
Given the size of the test sets I employ, this is almost certainly the case.

\subsection{Communication}

In the new algorithm, the data type of the flow directions and labels is fixed
at 1~byte/cell and 4~bytes/cell, respectively. The data type and, therefore,
size, of the elevations may change with the input data; call it $E$~bytes.
Disregarding data structure overhead, the new algorithm needs to pass the flow
directions, labels, and elevations of each tile's $4\sqrt{n}$ edge cells to the
producer at a cost of $(4\sqrt{n})(5+E)$ bytes. In addition each tile sends its
spillover graph. The spillover graph stores the minimal elevation of each
watershed's meeting point; therefore, each meeting point requires two labels and
one elevation. Since there are at most $4\sqrt{n}$ watersheds, if they all meet
this costs $(4\sqrt{n})(8+E)$ bytes; however, in practice this is an over-
estimate, as shown in Table~\ref{tbl:results}. In turn, for each tile the
producer passes back a mapping of each label to an elevation offset at a cost of
$(4\sqrt{n})(4+E)$. Therefore, the total communication cost is approximately
$(4\sqrt{n})(3E+17)$ per tile.

Previous parallel implementations have exchanged edge elevation information
between adjacent cores after each iteration of their algorithms. For a tiled
dataset the cost between two cores is $(2\sqrt{n})E$ bytes per iteration.
Therefore, the cost of communication between two cores in a previous algorithm
surpasses the cost of communication between a producer and a single consumer in
the new algorithm after $(6+34/E)$ iterations. Since, typically, $E>10^3$, this
is essentially six.

If the number of cores used by the new algorithm is $P$, then the communication
costs between any two cores of an iterative algorithm should surpass the cost of
all of the consumers communicating with a single producer in the new algorithm
after about $6P$ iterations. For the configuration used here, $P=48$, so this
number is about $288$, which is small in comparison to the size of the datasets
and, therefore, is likely to be exceeded.

%cat Array2D.hpp common.hpp main.cpp Zhou2015pf.hpp Barnes2014pf.hpp | grep -v "^\s*//" | grep -v "^\s*$"  |wc

\section{Empirical Tests}

\label{sec:tests}

I have implemented the algorithm described above in \texttt{C++11} using MPI for
communication, the Geospatial Data Abstraction Library (GDAL)~\citep{GDAL} to
read and write data, and Boost Iostreams to handle compression for the
\textsc{cacheC} strategy. Tests were performed using Intel MPI v5.1; the code is
also known to work with OpenMPI v1.10.2. There are 1505 lines of code and 203
lines of comments.\todo{Check again before publishing} Since the algorithm does
not rely on details of the communication, implementing the algorithm with Spark
or MapReduce or would be straight-forward. The code can be acquired from
\url{https ://github.com/r-barnes/Barnes2016-ParallelPriorityFlood}.

\newcolumntype{R}[2]{%
    >{\adjustbox{angle=#1,lap=\width-(#2)}\bgroup}%
    l%
    <{\egroup}%
}
\newcommand*\rot{\multicolumn{1}{R{45}{1em}}}% no optional argument here, please!

\begin{table*}
\footnotesize
\centering
\begin{tabular}{l l l l l lll S[table-format=3.2] S[table-format=3.2] l S[table-format=3.2]}
{DEM}         & {Resolution}     & {Tiles}     & {Cells/Tile}     & {Tile Size}    & {Total Size}     & {Cells}       \\ \hline
SRTM Resampled & 10\,m            & 14297       & 10803$^2$        & 233\,MB        & 3.34\,TB         & $1.7\cdot10^{12}$ \\
SRTM Global    & 30\,m            & 14297       & 3601$^2$         & 26\,MB         & 371\,GB          & $1.9\cdot10^{11}$  \\
NED            & 10\,m            & 1023        & 10812$^2$        & 468\,MB        & 478\,GB          & $1.2\cdot10^{11}$  \\
PAMAP North    & 1\,m             & 6666        & 3125$^2$         & 39\,MB         & 260\,GB          & $6.5\cdot10^{10}$  \\
PAMAP South    & 1\,m             & 6723        & 3125$^2$         & 39\,MB         & 263\,GB          & $6.6\cdot10^{10}$  \\
SRTM Region 1  & 30\,m            & 164         & 3601$^2$         & 25.9\,MB       & 4.3\,GB          & $2.1\cdot10^{ 9}$  \\
SRTM Region 2  & 30\,m            & 161         & 3601$^2$         & 25.9\,MB       & 4.2\,GB          & $2.1\cdot10^{ 9}$  \\ \hline
%SRTM Region 3  & 30\,m            & 150         & 3601$^2$         & 25.9\,MB       & 3.9\,GB          & $1.9\cdot10^{ 9}$ \\
%SRTM Region 4  & 30\,m            & 156         & 3601$^2$         & 25.9\,MB       & 4.0\,GB          & $2.0\cdot10^{ 9}$ \\
%SRTM Region 5  & 30\,m            & 165         & 3601$^2$         & 25.9\,MB       & 4.3\,GB          & $2.1\cdot10^{ 9}$ \\
%SRTM Region 6  & 30\,m            & 169         & 3601$^2$         & 25.9\,MB       & 4.4\,GB          & $2.2\cdot10^{ 9}$ \\
%SRTM Region 7  & 30\,m            & 152         & 3601$^2$         & 25.9\,MB       & 3.9\,GB          & $2.0\cdot10^{ 9}$ \\
\end{tabular}
\caption{Datasets employed for testing the new algorithm.
\textbf{Tiles} indicates the number of tiles the DEM was divided into by its
provider. \textbf{Tile Size} indicates how much uncompressed space it would take
to store the number of cells in the tile, given its data type (cell count times
data type size). \textbf{Total Size} indicates how much space it would take to
store all of the tiles in the dataset.
\label{tbl:my_datasets}}
\end{table*}

To demonstrate the scalability and speed of the algorithm, I tested it on
several large DEMs, including one \textit{rather} large one, as shown in
\autoref{tbl:my_datasets}. All of these DEMs came pre-divided into equally-sized
tiles by their providers; I used these existing tile structures in most of my
tests; however, my implementation of the algorithm can also break a monolithic
DEM into tiles suitable for processing, and this is also done.

The DEMs tested include
\begin{itemize}
\item PAMAP\footnote{\url{ftp://pamap.pasda.psu.edu/pamap_LiDAR/cycle1/DEM/}}: A LiDAR
DEM covering the entire state of Pennsylvania. The data is available as 13,918
tiles divided into a north section and a south section. These sections are
projected differently and, therefore, the two are considered independently here.
%At a resolution of 0.98\,m (3.2\,ft).%The PAMAP LiDAR elevation data was
%collected from 2006--2008 with 1.4\,m average point spacing and a vertical
%accuracy at least 18.5\,cm (RMSE) in open areas.

\item NED\footnote{\url{ftp://rockyftp.cr.usgs.gov/vdelivery/Datasets/Staged/Elevation/13/IMG/}}:
National Elevation Dataset 10\,m data. Higher resolution 3\,m and 1\,m data are
available, but only in patches, whereas 10\,m data are available for the entire
conterminous United States, Hawaii, and parts of Alaska. The entire 10\,m NED
DEM is considered here as a single unit. Although islands are present in the
DEM, the algorithm implicitly handles these without an issue.
%The mean absolute vertical accuracy of the dataset is
%$<$1\,m.~\citep{Gesch2014} 

\item SRTM:
Shuttle Radar Topography Mission (SRTM) 30\,m DEM. This 30\,m data covers 80\%
of Earth's landmass between 56$^\circ$S and 60$^\circ$N. The data was
originally available as several regions covering North
America\footnote{\url{http://dds.cr.usgs.gov/srtm/version2_1/SRTM1/}}, which
are considered separately here; more recently, global
data\footnote{\url{http://e4ftl01.cr.usgs.gov/SRTM/SRTMGL1.003/2000.02.11/}}
has been released. The global data is considered as a single unit here. Since the surfaces of oceans and the like are topographically uninteresting tiles which would contain only oceans are not present in the dataset.

\item There are not many datasets available which are large enough to tax the
algorithm described here, so I resampled the SRTM global data to three times
its original resolution (30\,m to 10\,m). This resulted in a
\textit{rather} large DEM which is henceforth called SRTM-RG.
\end{itemize}

Further details on acquiring the aforementioned datasets are available with the
source code.

%The data was collected
%in February 2000 by the Space Shuttle Endeavour and mapped Earth's topography
%between 56$^\circ$S and 60$^\circ$N (80\% of the Earth's landmass) over an
%eleven day period using an imaging radar. Although 30\,m data was released for
%several regions of the United States shortly thereafter, only 90\,m data was
%released for other parts of the world. Beginning in 2014, the entire dataset was
%gradually released at 30\,m. The data have an absolute accuracy of $\le$16\,m.
%Here, each U.S.\ region is considered independently and the entire global
%dataset is considered as a single unit.

Tests were run on the Comet machine of the Extreme Science and Engineering
Discovery Environment (XSEDE)~\citep{xsede}. Each node of the machine has
2.5\,GHz Intel Xeon E5-2680v3 processors with 24 cores per node, 128\,GB of
DDR4 DRAM, and a 320\,GB of local SSD storage. Nodes are connected with 56\,Gbps FDR InfiniBand. Data were held in
Oasis: a 200\,GB/s distributed disk Lustre filesystem. Code was compiled using
GNU g++ 4.9.2. Although intermediate products could be stored in nodes' local
SSD burst memory, I do not do so here in order to subject the algorithm to a
more antagonistic environment.

Five tests were run. For the first four tests, the algorithm was run using the
\textsc{evict} strategy to simulate a minimal-resource environment. The fifth
test relaxed this and tested the algorithm in all modes.

The first test ran the algorithm on two nodes (48 cores) for each of the
datasets listed in \autoref{tbl:my_datasets} using the full dataset and all of
the available cores. The result is shown in \autoref{tbl:results}.

All of the datasets contain islands of data surrounded by empty tiles, or have
irregular boundaries. Therefore, in order to test scaling, the largest square
subset of contiguous tiles was identified in each dataset. The resulting subsets were 44\,x\,44 (PAMAP North and South), 39\,x\,39 (SRTM Global), 19\,x\,19 (NED), 11\,x\,11 (SRTM Region 1 and 2).

\begin{figure*}
\centering
\begin{tabular}{cc}
\includegraphics[width=0.45\textwidth]{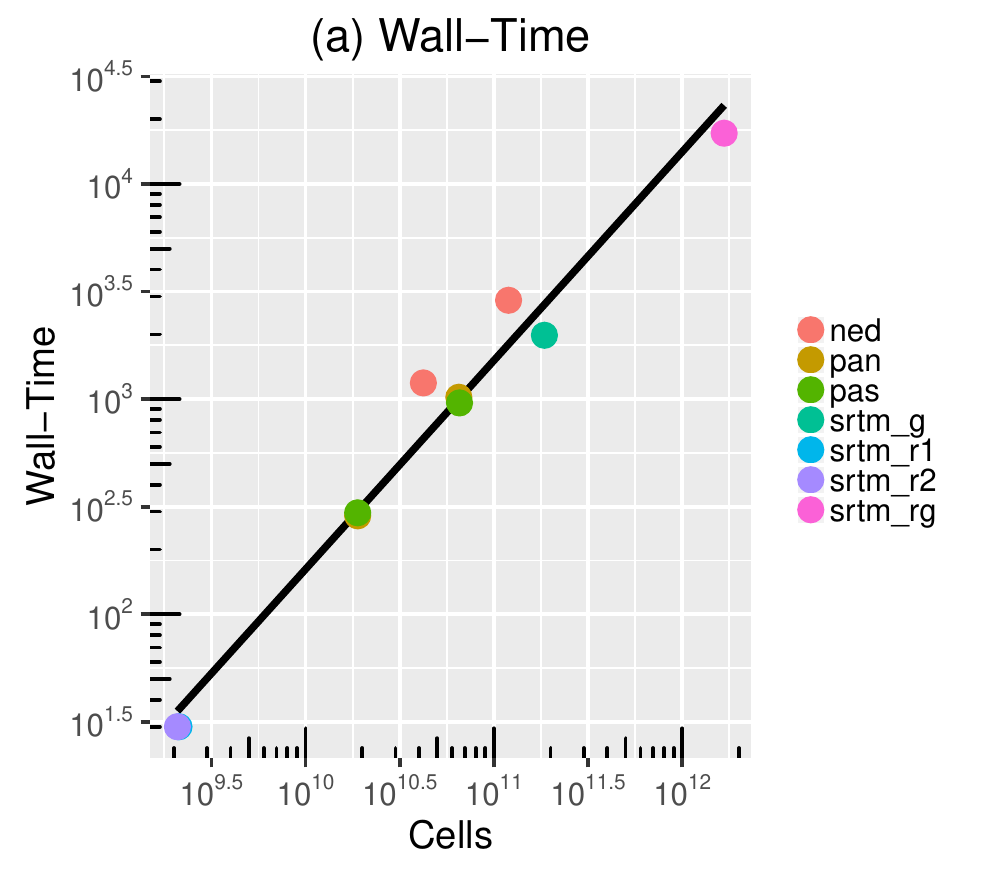} &
\includegraphics[width=0.45\textwidth]{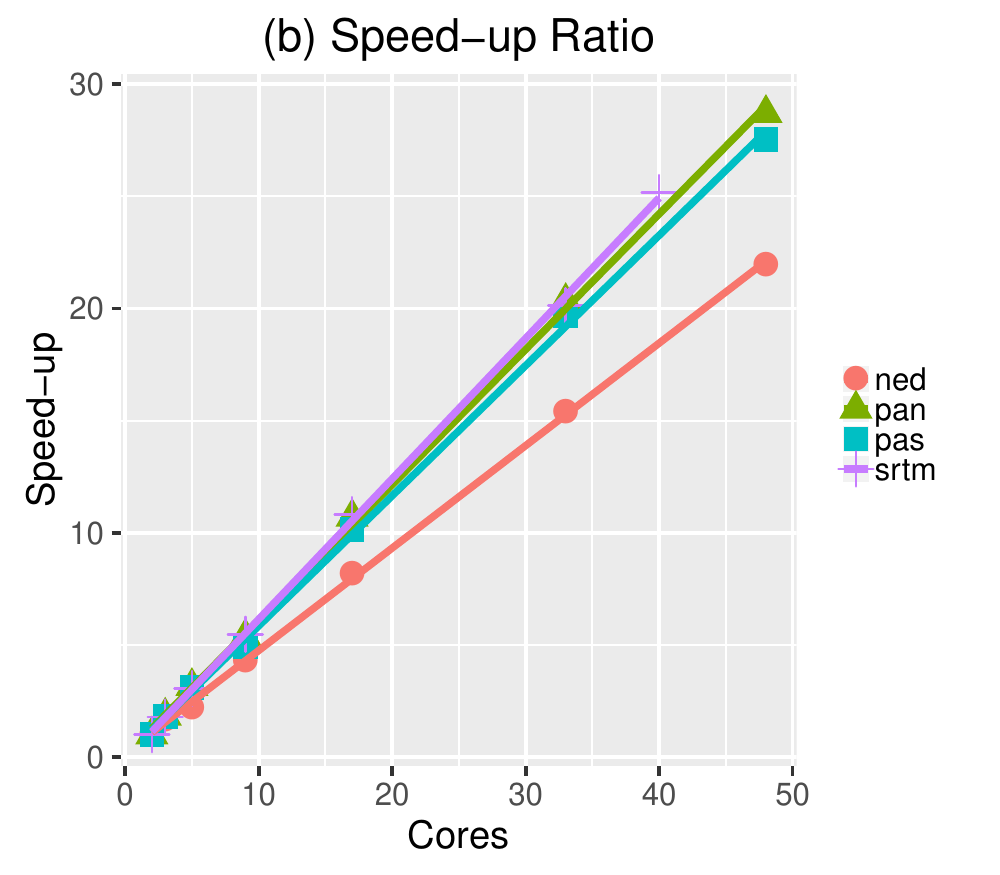} \\
\includegraphics[width=0.45\textwidth]{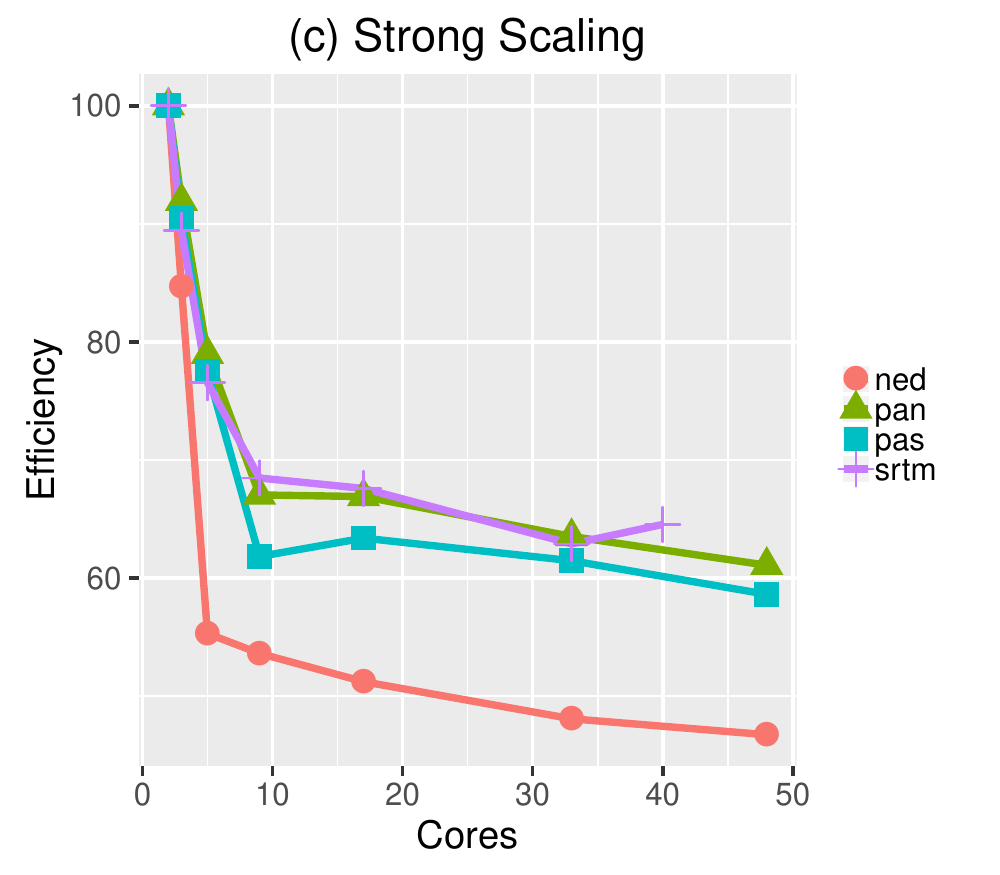} &
\includegraphics[width=0.45\textwidth]{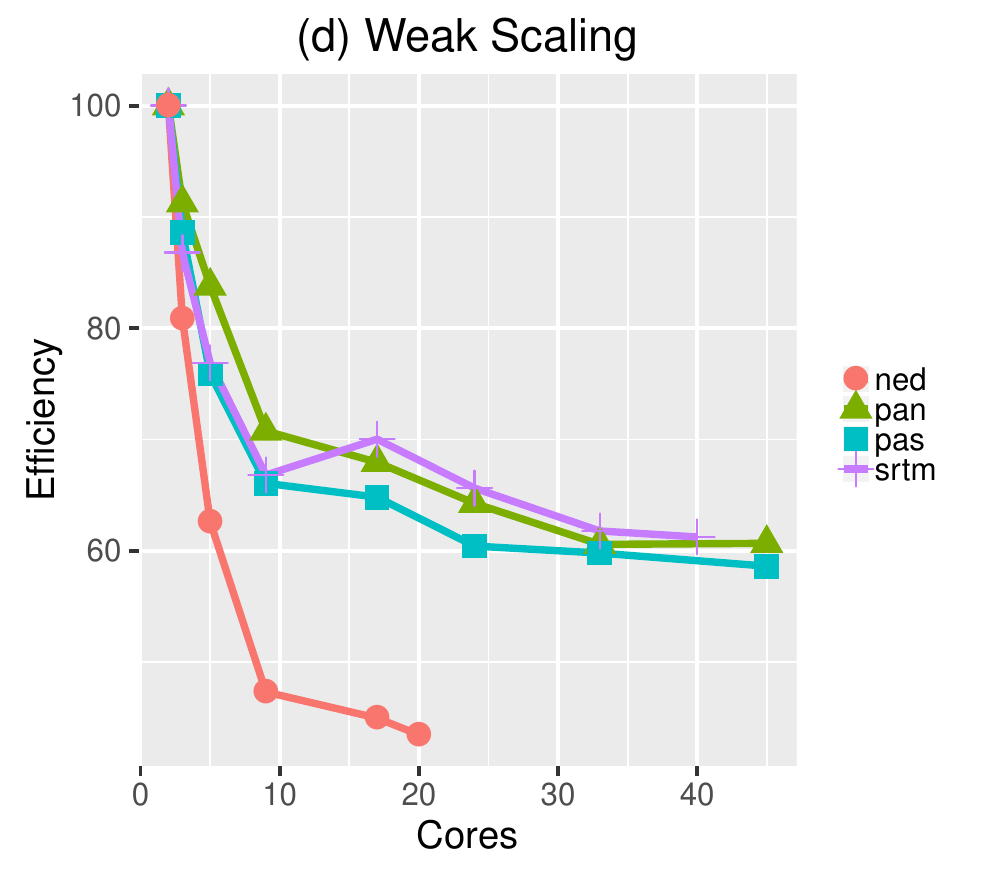}
\end{tabular}
\caption{Results. Let $N$ be the number of cores used, $t_1$ be the time taken by one core to perform one work unit, and $t_N$ be the time taken by $N$ cores to perform the job. The speed-up ratio is given as $\frac{t_1}{t_N}$ where the job size is unchanged. Strong scaling is given by $\frac{t_1}{N t_N}$ where the job size is unchanged. Weak scaling is given by $\frac{t_1}{t_N}$ where the job size is increased proportionally to $N$. \textbf{(a)} includes more than one point per dataset: the 48-core strong-scaling results have been included here to flesh out the trendline. \label{fig:result_graphs}}
\end{figure*}

The second and third tests were performed on these contiguous square subsets.
Strong scaling efficiency is a metric of an implementation's ability to solve a
problem faster by using more resources. To test this, increasing numbers of
cores (up to 48) were used on the full square subsets. Weak scaling efficiency
is a metric of an implementation's ability to solve proportionately larger
problems in the same time using proportionately more resources. To test this,
one core was used to process one row of each square subset, two cores for two
rows, and so on. The results are shown in \autoref{fig:result_graphs}.

In a fourth test, a comparison was made against the work of both \citet{Wallis2009depressions} (TauDEM\footnote{\texttt{e19dc083e}, master, \url{https://github.com/dtarb/TauDEM}}) and \citet{Gomes2012} (EMFlow\footnote{\texttt{0ca9e0ef0}, master, \url{https://github.com/guipenaufv/EMFlow}}). To handle the input limitations of EMFlow, a 40,000\,x\,40,000 single-file DEM was constructed by merging SRTM Region 2 data. All code was compiled using GNU g++ 4.9.2 with optimizations enabled. \texttt{usr/bin/time} and \texttt{mpiP}\footnote{\url{http://mpip.sourceforge.net}} were used to measure memory usage as well as communication times and loads. Both attach to programs at runtime, eliminating the need for modification.

TauDEM would not process the test dataset with only 2 cores, so a direct comparison with either EMFlow or the new algorithm in this configuration was not possible. Therefore, TauDEM and the new algorithm were compared using 48 cores distributed over 2 nodes, similar to all of the above tests. EMFlow is single-threaded and so was compared against the new algorithm using varying numbers of cores.

In the fifth test, the algorithm's various operating strategies (\textsc{evict}, \textsc{cache}, \textsc{cacheC}, and \textsc{retain}) were compared on a single node using the SRTM regional datasets. The \textsc{cache} and \textsc{cacheC} strategies utilized the node's local SSD for increased performance. The results are shown in \autoref{tbl:strategy_comparison}.

%TODO:  In testing, employing (b) resulted in a 13.7\% increase in
%time-to-completion versus (c) for the PAMAP North test described below.

\begin{table*}
\footnotesize
\centering
\begin{tabular}{l llll lll l l l l}
                & \rot{Time}       & \rot{Sec/$10^9$ cells}  & \rot{\% I/O} & \rot{All Time}       & \rot{Prod.\ Calc}    & \rot{Labels}    & \rot{Sent}        & \rot{Received}    & \rot{Tx/Tile}         & \rot{Cons.\ VmHWM}         & \rot{Prod.\ VmPeak} \\                  
DEM            & Min              &                         &  \%          & Hrs                  & Sec                  &                 & MB                & MB                & KB                    & MB                         & MB                  \\ \hline
SRTM Resampled  & 287              & 10                      &  11          & 223                  & 84                   & 21,625,210      & 50                & 4,109             & 291                   & 1,307                      & 12,236              \\
SRTM Global     & 33               & 11                      &  8           & 25.5                 & 37                   & 11,478,908      & 29                & 1,452             & 104                   & 209                        & 6,011               \\
NED             & 48               & 25                      &  4           & 37.1                 & 6                    & 1,451,911       & 6                 & 380               & 377                   & 1,725                      & 1,295               \\
PAMAP North     & 17               & 15                      &  9           & 12.8                 & 12                   & 2,384,615       & 12                & 717               & 109                   & 234                        & 1,943               \\
PAMAP South     & 16               & 15                      &  9           & 12.5                 & 10                   & 1,720,776       & 10                & 709               & 107                   & 233                        & 1,703               \\
SRTM Region 1   & 0.5              & 14                      &  3           & 0.32                 & 0.3                  & 95,106          & 0.3               & 16                & 99                    & 184                        & 478                 \\
SRTM Region 2   & 0.5              & 14                      &  3           & 0.31                 & 0.4                  & 139,472         & 0.4               & 17                & 105                   & 162                        & 494                 \\ \hline
%SRTM Region 3   & 0.37             & 6.17                    &  5.4        & 0.20                 &                      &                 &                   &                   &                       &                            &                     \\
%SRTM Region 4   & 0.46             & 7.74                    &  4.6        & 0.26                 &                      &                 &                   &                   &                       &                            &                     \\
%SRTM Region 5   & 0.44             & 7.09                    &  5.1        & 0.25                 &                      &                 &                   &                   &                       &                            &                     \\
%SRTM Region 6   & 0.37             & 6.09                    &  5.6        & 0.22                 &                      &                 &                   &                   &                       &                            &                     \\
%SRTM Region 7   & 0.25             & 3.64                    & 11.6        & 0.12                 &                      &                 &                   &                   &                       &                            &                     \\
\end{tabular}
\caption{Results.
\textbf{Time} is the time-to-completion (aka wall-time) of the
algorithm. \textbf{Sec/$10^9$ cells} indicates how many wall-time seconds it
took the algorithm to process a billion cells on each dataset. \textbf{All Time}
indicates the sum of the processing and I/O time of every CPU core used by the
algorithm; this is the unit supercomputing centers charge by. \textbf{\% I/O}
indicates what percentage of the All Time value was spent on reading and writing
data. \textbf{Prod.\ Calc} is the amount of time the producer spent calculating
the global solution. \textbf{Labels} is the number of unique, global labels
required. \textbf{Sent} is the amount of data sent by the producer.
\textbf{Received} is the amount of data received by the producer.
\textbf{Tx/Tile} is the sum of the data received and sent divided by the number
of tiles in the dataset. \textbf{Cons.\ VmHWM} is the virtual memory ``high
water mark" used by one of the consumers to store its data, as determined by the
Linux kernel. \textbf{Prod.\ VmPeak} is the peak virtual memory used by the
producer to store its data and the shared libraries it uses, as determined by
the Linux kernel.
\label{tbl:results}}
\end{table*}

\begin{table}
\footnotesize
\centering
\begin{tabular}{lcccc}
DEM           & \textsc{evict} & \textsc{cache} & \textsc{cacheC} & \textsc{retain} \\ \hline
SRTM Region 1 & 60 & 81 (1.4\,x)  & 51 (0.8\,x) & 34 (0.6\,x) \\
SRTM Region 2 & 56 & 80 (1.4\,x)  & 50 (0.9\,x) & 32 (0.6\,x) \\
%SRTM Region 3 & 46 & 69 (1.5\,x)  & 45 (1.0\,x)   & 26 (0.6\,x) \\
%SRTM Region 4 & 63 & 96 (1.5\,x)  & 52 (0.8\,x) & 34 (0.5\,x) \\
%SRTM Region 5 & 55 & 95 (1.7\,x)  & 47 (0.9\,x) & 34 (0.6\,x) \\
%SRTM Region 6 & 55 & 82 (1.5\,x)  & 47 (0.9\,x) & 32 (0.6\,x) \\
%SRTM Region 7 & 35 & 76 (2.2\,x)  & 28 (0.8\,x) & 20 (0.6\,x) \\ \hline
\end{tabular}
\caption{Timing results in seconds, and speed-up factors versus \textsc{evict}, for different caching strategies. \label{tbl:strategy_comparison}}
\end{table}

%Variation in this value is driven in part by different data formats.

\section{Results \& Discussion}
\label{sec:results}

\subsection{Comparisons}

In \textsection\ref{sec:alt_algs} I argue that the new algorithm should scale better than existing algorithms because it has lower time complexities, can use multiple cores, and has fixed I/O and communication requirements. The results of my tests support this.

%New alg, one consumer, 400x400
%VMPeak+1*VmHWM
%614368+1*60124

%New alg, one consumer, 4000x4000
%VMPeak+1*VmHWM
%241728+1*208716

%New alg, five consumers, 4000x4000
%VMPeak+5*VmHWM
%254684+5*193476

EMFlow running with a maximum of 2\,GB RAM and tiles of 400\,x\,400 cells (the settings
discussed by \citet{Gomes2012}) had 494\,s wall-time and used 1.8\,GB RAM. The
new algorithm running with one consumer and 400\,x\,400 tiles had 1,015\,s wall-time (2\,x more) and used 674\,MB RAM (2.7\,x less).

I tested the effect of larger tile sizes by running EMFlow with 4,000\,x\,4,000
tiles. This gave 2,957\,s wall-time (6.0\,x more versus 400\,x\,400 tiles) and used 1.8\,GB RAM. Compared to this, the new algorithm with one consumer and 4,000\,x\,4,000 tiles gave a wall-time of 583\,s (1.2\,x more) and used 450\,MB RAM (4\,x less). Running the new algorithm with five consumers and 4,000\,x\,4,000 tiles gave a wall-time of 170\,s (2.9\,x less) and used 1.2\,GB RAM (1.5\,x less).

It is notable that EMFlow with small tiles and a single processor runs just
1.2\,x faster than the new algorithm with a single consumer and large tiles.
EMFlow uses an $O(n)$ integer Priority-Flood based on hierarchical queues
whereas my implementation of the new algorithm uses a $O(n\log n)$ variant
suitable for any data type. In this case, it seems that the cost of generality
is small. It is also notable that when the new algorithm uses five consumers, it runs significantly faster than EMFlow while using less RAM.

%taudem_test.2538682.comet-18-38.out
%RAM
%(364488 + 364044 + 364044 + 384996 + 585132 + 867876 + 881852 + 858296 + 861856 + 884256 + 783680 + 842844 + 835096 + 785032 + 849748 + 813780 + 881672 + 799364 + 819732 + 819660 + 823120 + 822568 + 843912 + 844008 + 863308 + 853996 + 852456 + 862500 + 868688 + 848924 + 837984)/31*48
%Time
%(143.85+143.90+143.72+143.91+143.81+143.77+143.23+143.80+142.72+143.92+144.03+143.80+143.90+142.79+143.88+143.50+144.02+143.95+143.95+143.64+143.81+143.33+143.95+142.06+143.61)/25
%Total message data transmitted in bytes
%(4.21e+08)+(4.21e+08)+(1.5e+07)+(1.5e+07)+(1.5e+07)+(1.06e+04)+(1.92e+03)+(380)+(188)
%Comm time in seconds:
%((2.83e+06)+(2.12e+06)+(3.91e+05)+(3.41e+05)+(1.22e+04)+(1.07e+04)+(1.01e+04)+(9.03e+03)+(4.43e+03)+(203)+(119)+(106)+(102)+(57.9)+(54.6)+(16.8)+(16.6)+(14)+(11.1)+(6.96))/1000

%taudem_dpf_test.2538539.comet-02-37.out
%Total RAM:
%VmPeak+47*VmHWM
%(484048+47*97672)/1000 - Kilobytes -> Megabytes

On the 40,000\,x\,40,000 test set, TauDEM had 144\,s wall-time, transmitted
887\,MB, used 5,729\,s for communication, and took 37\,GB RAM. The new algorithm (running with a tile size of 4,000\,x\,4,000) had 23\,s wall-time (6.3\,x faster), transmitted 46\,MB (19\,x less), used 82\,s for communication (70\,x less), and took 5.1\,GB RAM (7.3\,x less). Communication time is greater than wall-time because it is a summation across many cores. The foregoing confirms many of the predictions made in \textsection\ref{sec:complexity}.

\subsection{Flexible Operation}

The above demonstrates that the algorithm can
leverage many-core systems, but also operate well with much more limited
resources. \autoref{tbl:results} provides further confirmation of this. VmPeak
shows the maximum RAM used by the producer to hold both its data and the shared
libraries used by the program and VmHWM shows the maximum RAM used by a
consumer. Since the producer and consumers trade off operation, they do not
contend for computational resources. Therefore, the memory required to process a DEM using only one consumer is approximately the sum of VmHWM and VmPeak: 13.5\,GB for a 3.34\,TB dataset in the largest case. 6.5\,GB would be sufficient to process any of the datasets mentioned in \autoref{tbl:whatismassive}. The time required for such an operation is given by the ``All Time" column of \autoref{tbl:results}, since the time required for calculations by the producer is negligible ($<84$\,s in the largest case).

As \autoref{tbl:strategy_comparison} shows, running the algorithm's various strategies on the SRTM regional data provides further evidence of the algorithm's flexibility. While the \textsc{cache} strategy does not seem to provide a performance advantage, the \textsc{cacheC} strategy saves several seconds of processing time. On larger datasets, this could make a noticeable difference. Clearly, when resources are available, utilizing the \textsc{retain} strategy is worthwhile.

% Region_01  @cache       81.3932
% Region_01  @cache       51.4323
% Region_01  @offloadall  59.5196
% Region_01  @retainall   34.3737
% Region_02  @cache       79.7439
% Region_02  @cache       49.8111
% Region_02  @offloadall  56.4321
% Region_02  @retainall   32.3018
% Region_03  @cache       68.8794
% Region_03  @cache       45.1198
% Region_03  @offloadall  46.3286
% Region_03  @retainall   26.2113
% Region_04  @cache       95.8883
% Region_04  @cache       51.6622
% Region_04  @offloadall  63.2309
% Region_04  @retainall   33.9256
% Region_05  @cache       95.195
% Region_05  @cache       47.4837
% Region_05  @offloadall  54.6771
% Region_05  @retainall   33.6985
% Region_06  @cache       81.6882
% Region_06  @cache       46.7226
% Region_06  @offloadall  54.8944
% Region_06  @retainall   32.3672
% Region_07  @cache       76.1075
% Region_07  @cache       27.6305
% Region_07  @offloadall  34.7648
% Region_07  @retainall   20.3338

%If fewer resources are available, the algorithm the tiles can be divided
%between a limited number of cores. In the most extreme case, tiles can be
%processed one at a time. This means the architecture of the machine the
%algorithm runs on does not affect its ability to run: large datasets can still
%be processed efficiently. The algorithm also offers several memory retention
%strategies (mentioned above) to help it adapt to different environments. This
%contrasts with previous algorithms which, in many cases, cannot run unless they
%are matched to an appropriate architecture.

\subsection{Scaling}

In \textsection\ref{sec:complexity}, I argued that the
algorithm should scale linearly with the number cells for a fixed tile size.
\autoref{fig:result_graphs}a confirms this: a linear fit to the log-log plot
has a slope of 0.97 (R$^2=0.99$) across datasets whose sizes differ by three
orders of magnitude. The NED data points are likely higher than the trend line
due to their larger tile sizes.

Figures \ref{fig:result_graphs}c and \ref{fig:result_graphs}d show sustained
efficiencies of 60\% on up to 48 cores distributed across two nodes for the
datasets with smaller tiles. The larger tiles of the NED result in lower scaling
efficiencies of 55\%, but this too remains nearly constant as the number of
cores increases. As a result, as the number of cores increases, the speed-up
ratio shown in Figure~\ref{fig:result_graphs}b is approximately linear with an
average slope of 0.56 across all datasets. This contrasts with the results of
\citet{Yildirim2015} whose implementation quickly reached diminishing returns
(see their Figure 7).

Additionally, note that the 21,625,210 unique watershed labels required for the
largest dataset fall well below the 4,294,967,295 threshold of an unsigned
32-bit integer (at which point a larger data type would be required).

\subsection{Larger datasets}

Can even larger, perhaps even \textit{unusually} large, datasets be
used? Yes. No fundamental limit prevents the algorithm from scaling to even
larger datasets than those tested here. As Figure~\ref{fig:result_graphs} shows,
the algorithm's time complexity is linear and it scales well across large
numbers of cores. Additionally, the processing time required by the producer is
negligible in comparison to the total, and the per tile communication
requirements are low. Although the 13.5\,GB RAM and 9.3 compute-days required for
the SRTM-RG dataset are near the limits of a high-spec laptop, they are well
within what a server or supercomputer is capable of.

A more complex implementation could reduce the producer's requirements by
performing partial computation of the global solution as tiles return their
data. For clarity, I have opted to build a simpler implementation which stores
all of the tiles' returned data in memory prior to calculating the global
solution. This is why the producer requires such a large amount of memory.

\subsection{Speed improvements} The algorithm can run faster. As discussed generally by
\citet{Hendriks2010} and in the context of Priority-Flood by
\citet{Barnes2014pf}, many priority-queue implementations are available and
some are much faster than others. In addition, $O(N)$ priority-queues such as
radix heaps and hierarchical queues are available for integer and
specially-formatted floating-point data. For my implementation, I have used the
general-purpose $O(N\log N)$ \texttt{C++ STL} priority-queue. While faster
implementations exist, the STL is general and well-tested, making it a safe
choice. The work of \citet{Zhou2015} also suggests that faster implementations of the serial Priority-Flood may be possible.

\subsection{Robustness}

The algorithm is robust in the face of crashes and other interruptions. The data each tile sends to the central node could be cached allowing the
algorithm to proceed without having to repeat work after a crash. Once the
central node has calculated a global solution, this solution can be cached and
distribution to tiles, along with output-generation, can continue after an
interruption. For simplicity, I have not yet included this capability in my own
implementation.

\subsection{Correctness}

I believe it is and hope that the foregoing
description and pseudocode will be sufficient for an interested reader to
convince themselves of this. But a seemingly convincing proof may be flawed.
Therefore, I have built an automated tester which performs correctness tests on
arbitrary inputs. This tester, along with several tests, is included in the
source code.

Of the works cited in Table~\ref{tbl:whatismassive}, none describe a correctness
testing methodology, though several~\citep{Do2011,Metz2011,Metz2010} compare the
results of stream network extraction between algorithms or other data sources.
Unfortunately, since this end result will differ by methodology it cannot be
used as an argument for algorithmic correctness.

In any test, a correct result must be established. While ArcGIS or GRASS could
be used for this, doing so would introduce a large and potentially expensive
dependency that could not be included with the source code. Therefore, I run a
simple implementation of the Priority-Flood on the entire DEM to establish
correct results. This algorithm is well-established in the field and its
implementation is simple enough that its correctness can be established by
inspection.~\citep{Barnes2014pf}

In testing, if a single file is given as input, an authoritative answer is
generated from the file as described above. The file is then subdivided into
tiles. A large number of different tile dimensions are tested to ensure that the
results of the new algorithm agree with the authoritative answer independent of
the tile dimension used. If a pre-tiled dataset is given as input, the tiles are
merged using \textsc{gdal} and treated as a single unit to generate an
authoritative answer. The algorithm is then run on the uncombined tiles. In all
cases, the algorithm is run with each of its memory retention strategies.
Running this suite of tests on a number of inputs did not show any deviation
from the authoritative answer, which is evidence of correctness. The source code
available with this paper includes this test suite.

\section{Coda}

A limitation of the algorithm presented here is that it only fills depressions;
often, though, flow accumulation is also desired. To obtain it, flow directions
must be calculated~\citep{Barnes2014pf,Barnes2014dd}; however, care is needed to
ensure that the methods used for doing so do not break the bounds on the number
of communication and I/O events established here. In future work, I will
describe how this problem can be overcome, and flow directions assigned.
Additionally, it may be possible to extend the techniques described here to
implement depression breaching in a manner similar to that described by
\citet{Lindsay2015}.

Once flow directions are assigned, \citet{Barnes2011} has provided a theoretical
description of an algorithm which permits the calculation of flow accumulation
using a fixed number of I/O and communication events per tile. In future work, I
will couple this algorithm with that presented here to construct a complete
package for processing \textit{rather} large DEMs. This work can likely be
extended to incorporate ideas from the ``flow algebra" described by
\citet{Tarboton2008} to form a very general approach for extracting hydrological
features and properties from DEMs.

In summary, prior depression-filling algorithms for large digital elevation
models required massive centralized RAM, suffered from unpredictable and slow
disk access when a virtual tile approach was used, or required large numbers of
nodes and communications when parallel processing was used. In contrast, the
present work has introduced a new algorithm which ensures fixed numbers of disk
accesses and communication events. This enables the efficient processing of
\textit{rather} large DEMs on both high- and low-resource machines.

Complete, well-commented source code, an associated makefile, and correctness
tests are available at
\url{https://github.com/r-barnes/Barnes2016-ParallelPriorityFlood}. This
algorithm is part of the RichDEM (\url{https://github.com/r-barnes/richdem})
terrain analysis suite, a collection of state of the art algorithms for
processing large DEMs quickly.

%TODO:
%Future work: \epsilon > 0

\section{Acknowledgments}
I gratefully acknowledge Clarence Lehman and Adam Clark who, at separate times,
provided the laptops upon which development of this algorithm took place.
Early-stage development utilized supercomputing time and data storage provided
by the University of Minnesota Supercomputing Institute. Empirical tests and
results were performed on XSEDE's Comet supercomputer~\citep{xsede}, which is
supported by National Science Foundation grant number ACI-1053575. Several
members of Berkeley's Energy \& Resources Group provided helpful feedback on
the paper, as did two anonymous reviewers.

My work was supported by the National Science Foundation's Graduate Research
Fellowship and by the Department of Energy's Computational Science Graduate
Fellowship (Grant No.\ DE-FG02-97ER25308).

\section{Bibliography}

{\footnotesize
  \bibliography{refs}   % expects file "myrefs.bib"
}

\end{document}